\documentclass[aps,11pt,prd,longbibliography]{revtex4-1}

\usepackage{feynmp-auto}
\usepackage{amsmath}
\usepackage[utf8]{inputenc}

\usepackage{bbm}
\usepackage{soul}	
\usepackage{xcolor}

\usepackage{epsfig}
\usepackage{color}
\usepackage{slashed}
\usepackage{comment}
\usepackage{epstopdf}
\usepackage{array}
\usepackage{booktabs}
\usepackage{mathrsfs}
\usepackage[euler]{textgreek}
\usepackage{amsmath}
\usepackage{amsthm}
\usepackage{amsfonts}
\usepackage{amssymb}
\usepackage{graphicx}
\graphicspath{ {Figures/} }
\usepackage[font={small}]{caption}
\usepackage{subcaption}
\usepackage{float}
\usepackage{multirow}
\usepackage[export]{adjustbox}
\usepackage[hang, flushmargin,bottom]{footmisc} 
\usepackage{hyperref}
\hypersetup{
     colorlinks   = true,
     citecolor    = blue
}
\usepackage{placeins}
\usepackage{enumitem}
\usepackage{slashed}
\usepackage{bbm}
\usepackage{appendix}

\usepackage{titlesec}
\titleformat{\chapter}[display]
  {\normalfont\LARGE\bfseries}
  {\chaptertitlename\ \thechapter}{5pt}{\LARGE}
  \titlespacing*{\chapter}{0pt}{-20pt}{35pt}
\usepackage{bigstrut}
\usepackage{pst-node}
\usepackage{pstricks}
\usepackage{physics}
\usepackage{mathtools}
\usepackage{setspace}
\usepackage{fancyhdr}
\usepackage[makeroom]{cancel}
\usepackage{feyn}
\newcommand{\be}{\begin{equation}}
\newcommand{\ee}{\end{equation}}
\newcommand{\bes}{\begin{equation*}}
\newcommand{\ees}{\end{equation*}}

\usepackage{hyperref}% http://ctan.org/pkg/hyperref
\hypersetup{%
  colorlinks = true,
  linkcolor  = black
}
\usepackage{soul}

\newcommand{\beq}{\begin{equation}}
\newcommand{\eeq}{\end{equation}}

\newcommand{\SU}{\,{\rm SU}}
\newcommand{\U}{\,{\rm U}}

\AtBeginDocument{%
    \newwrite\bibnotes
    \def\bibnotesext{Notes.bib}
    \immediate\openout\bibnotes=\jobname\bibnotesext
    \immediate\write\bibnotes{@CONTROL{REVTEX41Control}}
    \immediate\write\bibnotes{@CONTROL{%
    apsrev41Control,author="08",editor="1",pages="1",title="0",year="1"}}
     \if@filesw
     \immediate\write\@auxout{\string\citation{apsrev41Control}}%
    \fi
}%

\begin{document}
%%%%%%%%%%%%%%%%%%%%%%%%%%%%%%%%%%%%%%%%%%%%%%%%%%%%%%%%%%%%%%%%%%%%%%%%%%%%%%%%%%

%%%%%%%%%%%%%%%%%%%%%%%%%%%%%%%%%%%%%
\title{{\Large{\bf{On Anomaly-Free Dark Matter Models}}}}
\author{Pavel Fileviez P\'erez$^{1}$, Elliot Golias$^{1}$, Rui-Hao Li$^{1}$, Clara Murgui$^{2}$, Alexis D. Plascencia$^{1}$}
\affiliation{$^{1}$Physics Department and Center for Education and Research in Cosmology and Astrophysics (CERCA), 
Case Western Reserve University, Rockefeller Bldg. 2076 Adelbert Rd. Cleveland, OH 44106, USA \\
$^{2}$Departamento de F\'isica Te\'orica, IFIC, Universitat de Valencia-CSIC, 
E-46071, Valencia, Spain}
\email{pxf112@case.edu, ebg23@case.edu, rxl527@case.edu, clara.murgui15@gmail.com, alexis.plascencia@case.edu}
\date{\today}
\begin{abstract}
We investigate the predictions of anomaly-free dark matter models for direct and indirect detection experiments. We focus on gauge theories where the existence of a fermionic dark matter candidate is predicted by anomaly cancellation, its mass is defined by the new symmetry breaking scale, and its stability is guaranteed by a remnant symmetry after the breaking of the gauge symmetry. We find an upper bound on the symmetry breaking scale by applying the relic density and perturbative constraints. The anomaly-free property of the theories allows us to perform a full study of the gamma lines from dark matter annihilation. We investigate the correlation between predictions for final state radiation processes and gamma lines. Furthermore, we demonstrate that the latter can be distinguished from the continuum gamma ray spectrum.
\end{abstract}
\maketitle 

%\tableofcontents

%%%%%%%%%%%%%%%%%%%%%%%%%%%
\section{INTRODUCTION}
%%%%%%%%%%%%%%%%%%%%%%%%%%%%
\FloatBarrier

The nature of dark matter (DM) remains one of the most pressing issues in cosmology. In the past decades there has been a strong experimental effort 
to search for the particle nature of dark matter with no positive results so far. The lack of a discovery has pushed the scientific community to 
perform more general studies on dark matter models. The use of an effective field theory, in which all heavy degrees of freedom except for 
the dark matter particle are integrated out, covers a large set of dark matter models~\cite{Beltran:2008xg, Cao:2009uw,  Beltran:2010ww, Goodman:2010yf, Bai:2010hh}. 
Nonetheless, it has been shown that the effective field theory approach cannot lead to general conclusions, especially for collider studies where the energy scale can be larger than the cut-off scale.

An alternative framework for the study of different dark matter candidates consists of simplified models~\cite{Busoni:2013lha, Buchmueller:2013dya, Busoni:2014sya, Buchmueller:2014yoa, Buckley:2014fba, Harris:2014hga, Haisch:2015ioa}, in which the mediator between the dark matter and the Standard Model (SM) sector is also included in the particle spectrum. In this context, studies of dark matter can be done with only a few parameters 
and a systematic study can be performed on a large class of dark matter models. However, these simplified models do not come free of problems. 
They present issues of unitarity and gauge anomalies~\cite{Duerr:2013dza, Kahlhoefer:2015bea, Ismail:2016tod, Ellis:2017tkh, Cui:2017juz}. In particular, when considering indirect 
detection and the predictions for gamma lines from dark matter annihilation, a complete anomaly-free dark matter theory is needed, see 
for example~\cite{Duerr:2015wfa,FileviezPerez:2019rcj} for a detailed discussion.

There exist a large number of theories for dark matter; however, gauge theories can be very appealing due to the fact that the dark matter properties can be defined by a new gauge symmetry in nature. Then, it is possible to have different extensions of the Standard Model where one of the SM global symmetries is promoted to a local symmetry. 
For example, if we add three copies of right-handed neutrinos it is possible to have a theory based on local $B\!-\!L$. In this context, it is not possible to predict 
the existence of a dark matter candidate but extra fields are required to play this role, see for example the study in Ref.~\cite{FileviezPerez:2018toq}. 
The existence of a fermionic dark matter can be predicted in gauge theories when one of the fields needed for anomaly cancellation is in fact stable and has the right properties to describe the cold dark matter in the Universe. 
At the same time, the stability of the dark matter 
candidate can be a consequence of the new gauge symmetry breaking mechanism. The simplest theories we know where one can realize this 
idea are based on local baryon and lepton numbers~\cite{Duerr:2013dza,Perez:2014qfa}, for a recent review see \cite{Perez:2015rza}. The theories based on local baryon number could describe physics at the scale very close to the electroweak scale, and therefore, there is hope to test them in the near future.

In this article, we investigate the dark matter phenomenology in the context of simple theories where baryon number is a local gauge symmetry spontaneously broken at the low scale~\cite{Duerr:2013dza,Perez:2014qfa}. 
Our dark matter candidate is a Majorana fermion charged under the local $\U(1)_B$ gauge group, and hence, it has an axial coupling to the new gauge boson $Z_B$ present in the theory. This property gives a velocity-suppressed dark matter-nucleon interaction which means that bounds from direct detection experiments can be avoided. The simplest theories~\cite{FileviezPerez:2011pt,Duerr:2013dza,Perez:2014qfa} where baryon number is a local symmetry are very appealing extensions of the Standard Model where the spontaneous breaking of baryon number at the low scale can be understood, the stability of the proton is predicted at any level in perturbation theory, there is a good dark matter candidate in the theory, and there could be interesting mechanisms to explain the baryon asymmetry of the Universe. 

We perform a detailed study of the relic density, experimental constraints coming from the LHC and 
direct detection experiments. From the requirement of not overproducing dark matter combined with perturbativity, we find an upper bound on the symmetry breaking scale of the theory $\lesssim 28 $ TeV. 
Therefore, there is a hope to test these theories at the Large Hadron Collider or future colliders.  
We demonstrate that having a Majorana candidate in this type of theories allows the gamma lines to be visible. This is due to the fact that final state radiation processes are suppressed. The predictions for gamma lines are very important because gamma lines provide a clean and distinctive signature, and hence, they represent a smoking gun for the discovery of the dark matter. 

This article is organized as follows. In Section~\ref{sec:AnomalyFree}, we discuss the main features of anomaly-free dark matter models, while in Section~\ref{sec:SimplifiedModel}, we present a discussion of the simplified model of dark matter, compute the dark matter relic abundance and study LHC, direct detection and perturbativity constraints on the model. In Section~\ref{sec:GammaLines}, we show that a consistent anomaly-free theory leads to visible gamma lines in the photon spectrum of dark matter annihilation. We present our concluding remarks in Section~\ref{sec:Summary}.
%%%%%%%%%%%%%%%%%%%%%%%%%%%%%%%%%%%%%%
\section{ANOMALY CANCELLATION AND DARK MATTER CANDIDATES}
\label{sec:AnomalyFree}
%%%%%%%%%%%%%%%%%%%%%%%%%%%%%%%%%%%%%%
For simplicity, let us consider a simple Abelian gauge theory and take all the SM fermions to be charged under this new $\U(1)^{'}$, then, the existence of a dark matter candidate can be predicted from the requirement of anomaly cancellation. However, if the new symmetry is anomaly-free, for 
example $\U(1)_{L_i - L_j}$ or $\U(1)_{B-L}$ (with extra right-handed neutrinos), the existence of dark matter cannot be predicted because there is no need to add new degrees of freedom for anomaly cancellation.
In general, we can consider
$$Q_L ^i \sim (3,2,1/6,n_{Q_i}), \  u_R^i \sim (3,1, 2/3, n_{u_i}), \ d_R^i \sim (3,1,-1/3, n_{d_i}),$$ 
\vspace{-0.75cm}
$$\ell^i_L \sim (1,2,-1/2,n_{\ell_i}), \  {\rm{and}} \ e_R^i \sim (1,1,-1, n_{e_i}),$$
where each entry in the parenthesis corresponds to the quantum number for each multiplet under the gauge groups ($\SU(3)_C	$, $\SU(2)_L$, $\U(1)_Y$, $\U(1)_{B-L}$) and $i=1,2,3$ is the family index. Finding a simple solution to the anomaly cancellation conditions in agreement with all experimental bounds is a nontrivial exercise. After the discovery of the SM Higgs boson, it is possible to include extra fermions in a theory, but they should be vector-like under the SM gauge group, and if they are chiral under the new gauge symmetry they must acquire mass from the symmetry 
breaking  mechanism. It is possible to find different solutions for anomaly cancellation, but generically, these theories will have the following general features:
\begin{itemize}

\item There is an extra electrically neutral field, $\chi_{L( {\rm{or}} \, R)} \sim (1,1,0,n_\chi)$ which plays a role in the cancellation of anomalies, this field can be stable and hence a good cold dark matter candidate.

\item The mass of the DM candidate is determined by the new symmetry breaking scale.

\item The stability of the DM candidate is a natural consequence of the symmetry breaking. In the scenario where the DM is a Majorana fermion, the Abelian symmetry is broken to a $Z_2$ 
discrete symmetry, while in the Dirac case the stability is determined by an anomaly-free global symmetry.

\item The masses of the fermions required for anomaly cancellation have an upper bound defined by the new symmetry scale.

\item An upper bound on the symmetry breaking scale can be found by applying cosmological relic density constraints.

\end{itemize}

Perhaps, the simplest cases correspond to the theories where the baryon and/or lepton number are local symmetries~\cite{Duerr:2013dza,Perez:2014qfa}. In this article, for simplicity, we focus on the case where the new symmetry is the local baryon number, $\U(1)_B$,  because this theory can describe new physics at a scale very close to the electroweak scale in agreement with all experimental constraints. In our study, we will investigate the properties of the Majorana dark matter candidates 
since they are predicted in both theories proposed in Ref.~\cite{Duerr:2013dza,Perez:2014qfa}.

%%%%%%%%%%%%%%%%%%%%%%%%%%%%
\section{MAJORANA LEPTOPHOBIC DARK MATTER}
\label{sec:SimplifiedModel}
%%%%%%%%%%%%%%%%%%%%%%%%%%%%
In this article, we investigate the properties of a Majorana WIMP dark matter candidate in theories where baryon number is a local symmetry spontaneously broken at the low scale.
Here we discuss the main properties of our dark matter candidate. The relevant Lagrangian for our discussion is given by
\begin{eqnarray}
{\cal{L}} &\supset& - g_B n_\chi \bar{\chi} \gamma^\mu \gamma^5 \chi Z^B_\mu - g_B \bar{f} \left( n_V^f \gamma^\mu + n_A^f \gamma^\mu \gamma^5 \right) f Z^B_\mu 
- \lambda_i \bar{\chi} \chi h_i - \frac{1}{2} M_\chi {\chi}^T C \chi, 
\end{eqnarray}
where $\chi=\chi^C$ is a Majorana fermion. The first term defines the interaction between dark matter, $\chi$, and the new gauge boson $Z_B$, which mediates the new baryonic force. 
Here $g_B$ is the new gauge coupling, the fermions $f$ can be any SM quark or the new fermions needed for anomaly cancellation. The coefficients $\lambda_i$ define the interaction between the Higgses present in the theory and the DM candidate. Typically, one needs only two Higgses in these theories; the SM Higgs and the new Higgs needed to generate mass for the new gauge boson and fermions needed for anomaly cancellation. Therefore, the mechanism of spontaneous symmetry breaking needs to be applied in these theories, see Refs.~\cite{Duerr:2013dza,Perez:2014qfa} for details.

In order to study the dark matter phenomenology, the most relevant part of the Lagrangian in the theories proposed in Refs.~\cite{Duerr:2013dza,Perez:2014qfa} is given by
\begin{eqnarray} \label{eq:SimplifiedDM}
{\cal{L}} &\supset&  \frac{3}{4} g_B  \bar{\chi} \gamma^\mu \gamma^5 \chi Z^B_\mu - \frac{1}{3} g_B \bar{q} \gamma^\mu q Z^B_\mu  
+ \frac{ M_\chi}{2 v_B} \sin \theta_B \bar{\chi} \chi h_1 - \frac{M_\chi}{2 v_B} \cos \theta_B \bar{\chi} \chi h_2 - \frac{1}{2} M_\chi {\bar{\chi}} \chi, 
\end{eqnarray}
% if we assume 
where we have assumed the simple case where the DM is the Majorana fermion $\chi=\chi_L + (\chi_L)^C$ and neglect the mixing with the other fermions present in the theory for simplicity. The factor 3/4 is the baryon number of the dark matter divided by 2, the latter comes from the 1/2 factor in the projector. The gauge boson has vector coupling to quarks since $Z_B$ is associated to baryon number.
In this paper we will focus on this scenario since it is realized in both models in Refs.~\cite{Duerr:2013dza,Perez:2014qfa}. In the above equation, $\theta_B$ is the mixing angle between the SM Higgs and the new Higgs present in the theory to break the local baryon number symmetry, $S_B\sim (1,1,0,3)$. In the above equation $h_1$ and $h_2$ are the physical states present in the theory. 

The new symmetry breaking scale $v_B$ can be replaced by $v_B=M_{Z_B}/3 g_B$, where $M_{Z_B}$ is the mass of the new gauge boson. Henceforth, we set the SM Higgs boson mass to $M_{h_1}=125.09$ GeV and $v_H=246.22$ GeV. Then, this simplified model contains five free parameters,
\beq
M_\chi, \,\,\,\, M_{Z_{B}}, \,\,\,\, M_{h_2}, \,\,\,\, \theta_{B}, \,\,\,\, g_{B}.
\eeq
In Ref.~\cite{FileviezPerez:2018jmr} we have discussed in detail the experimental bounds on $M_{Z_B}$ and $\theta_B$. In the next section we will discuss the impact of the cosmological bounds and the possibility to find an upper bound on the symmetry breaking scale. The simplified model described in Eq.~\eqref{eq:SimplifiedDM} can arise from the following gauge-invariant Lagrangian,
%%%%%%%%%%%%%%%%%%%%%%%%%%%%%%%%%%%
\begin{align}
\mathcal{L} \supset &  \ i \overline{\chi}_L \gamma^\mu D_\mu \chi_L  - \left(  \frac{\lambda_\chi}{\sqrt{2} } \, \chi_L^T C \chi_L S_B^* + \rm{h.c.} \right),
\end{align}
%%%%%%%%%%%%%%%%%%%%%%%%%%%%%%%%%%%
where $D^\mu \chi_L=\partial^\mu \chi_L + i (3 g_{B}/2)  Z_{B}^\mu \chi_L$. Defining a Majorana field $\chi=\chi_L + (\chi_L)^C$ one can obtain the terms in Eq.~\eqref{eq:SimplifiedDM},
whereas the dark matter mass is given by
\beq
M_\chi=\lambda_{\chi} v_B. 
\eeq
The scalar potential is given by,
\begin{align}
V(H, S_{B}) = & -\mu_H^2 H^\dagger H - \mu_{B}^2 S_{B}^\dagger S_{B} + \lambda_H (H^\dagger H )^ 2  + \lambda_{B} (S_{B}^\dagger S_{B} )^2 + \lambda_{HB} (H^\dagger H) (S_{B}^\dagger S_{B}), 
\end{align}
where $H$ corresponds to the SM Higgs doublet and $S_{B}$ is charged under the $\U(1)_B$ group. In the zero temperature vacuum of the theory, both fields acquire a non-zero vacuum expectation value, and we can write 
\beq
S_{B}=\frac{1}{\sqrt{2}} \left( s_{B}+v_{B} \right), \hspace{6mm}  H = \frac{1}{\sqrt{2}}\begin{pmatrix}
  0 \\  h+v_H
 \end{pmatrix},
\eeq
where the Higgs doublet has been written in the unitary gauge. This leads to mixing among both scalars, and hence, the mass matrix needs to be diagonalized in order to find the physical states. The latter are given by,
\begin{align}
h_1 &= h \cos \theta_{B} - s_{B} \sin \theta_{B}, \\
h_2 &= s_{B} \cos\theta_{B} + h \sin \theta_{B},
\end{align}
where the scalar mixing angle can be written in terms of the scalar quartic couplings and the vevs,
\beq
\tan 2\theta_{B}  = \frac{\lambda_{HB} \, v_H v_{B}} {\lambda_{B} v_{B}^2  - \lambda_H v_H^2}.
\eeq

In order to perform this study we need to understand all perturbative bounds on the free parameters of the model.   
In Ref.~\cite{FileviezPerez:2018jmr} we have pointed out the need to impose the perturbative bounds on the scalar couplings in the Higgs sector,
\begin{align}
\lambda_H = & \frac{1}{2v_H^2} \left( M_{h_1}^2 \cos^2 \theta_{B} + M_{h_2}^2 \sin^2 \theta_{B} \right) \leq 4 \pi, \\
\lambda_{B} = & \frac{1}{2v_{B}^2} \left( M_{h_1}^2 \sin^2 \theta_{B} + M_{h_2}^2 \cos^2 \theta_{B} \right) \leq 4 \pi, \\
\lambda_{HB} = & \frac{1}{v_H v_{B}} \left( M_{h_2}^2  - M_{h_1}^2 \right) \sin \theta_{B} \cos \theta_{B}  \leq 4 \pi .
\end{align}
%\lambda_{B} = & \frac{1}{2 v_{B}^2 \cos^2 \theta_{B} } \left( M_{h_2}^2 - 2\lambda_H v_H^2 \sin^2 \theta_{B} - \lambda_{HB} \, v_H v_{B} \sin 2\theta_{B} \right),
To ensure vacuum stability of the scalar potential we impose
\begin{align}
\lambda_H, \lambda_{B} > 0 \hspace{5mm} {\rm and} \hspace{5mm} \lambda_{HB} > -2\sqrt{\lambda_H \lambda_{B}},
\end{align}
for more details see Ref.~\cite{FileviezPerez:2018jmr}. 
For consistency of our calculation, the gauge coupling $g_B$ must remain perturbative. If we revisit all the gauge interactions present in the theory, symbolically we have that $\bar{q} Z_B q$, $\bar{\chi} Z_B \chi$, and $Z_B Z_B S_B S_B$, 
we find that the strongest upper bound on $g_B$ is coming from the $Z_B Z_B S_B S_B$ interaction and it reads as $g_B \leq \sqrt{2 \pi}/3$. Following the notation above, the perturbative bound on the Yukawa coupling between the dark matter and the Higgses is $\lambda_\chi < 2 \sqrt{\pi}$.

%%%%%%%%%%%%%%%%%%%%%
\subsection{Relic Density}
\begin{figure}[tb]
\centering
\includegraphics[width=0.9\linewidth]{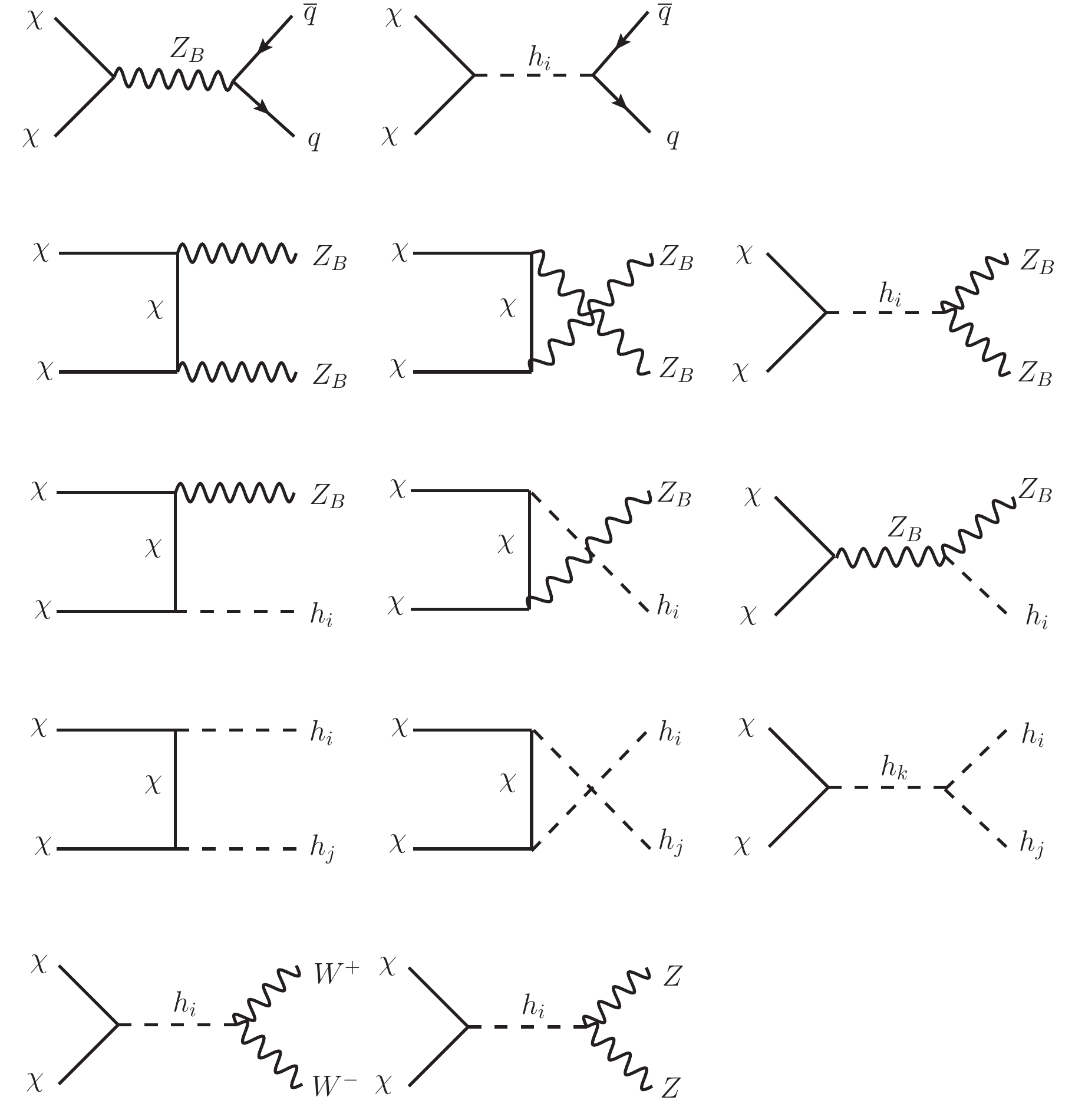} 
\caption{Feynman diagrams for the dark matter annihilation channels.}
\label{Diagrams}
\end{figure}

%%%%%%%%%%%%%%%%%%%%%  
The knowledge of all the details of the model allows us to investigate in detail the dark matter annihilation channels. 
In this context, the leptophobic dark matter can have the following annihilation channels,
$$\chi \chi \to \bar{q}q, \, Z_B Z_B, \, Z_B h_1, \, Z_B h_2, \, h_1 h_1, \, h_1 h_2, \, h_2 h_2, \, WW, \, ZZ.$$
The channels $Z_B h_1, \, h_1 h_1, \, h_1 h_2, \, WW$ and $ZZ$ are suppressed by the mixing angle $\theta_B$.  In Fig.~\ref{Diagrams} we show the Feynman graphs for each channel.
For our numerical study we implement the model in \texttt{LanHEP 3.2} \cite{Semenov:2014rea} and perform the calculation of $\Omega_{\rm DM} h^2$  using \texttt{MicrOMEGAs 5.0.6} \cite{Belanger:2018mqt}. Moreover, we perform an independent calculation in \texttt{Mathematica}. 

\begin{figure}[t]
\centering
\includegraphics[width=0.48\linewidth]{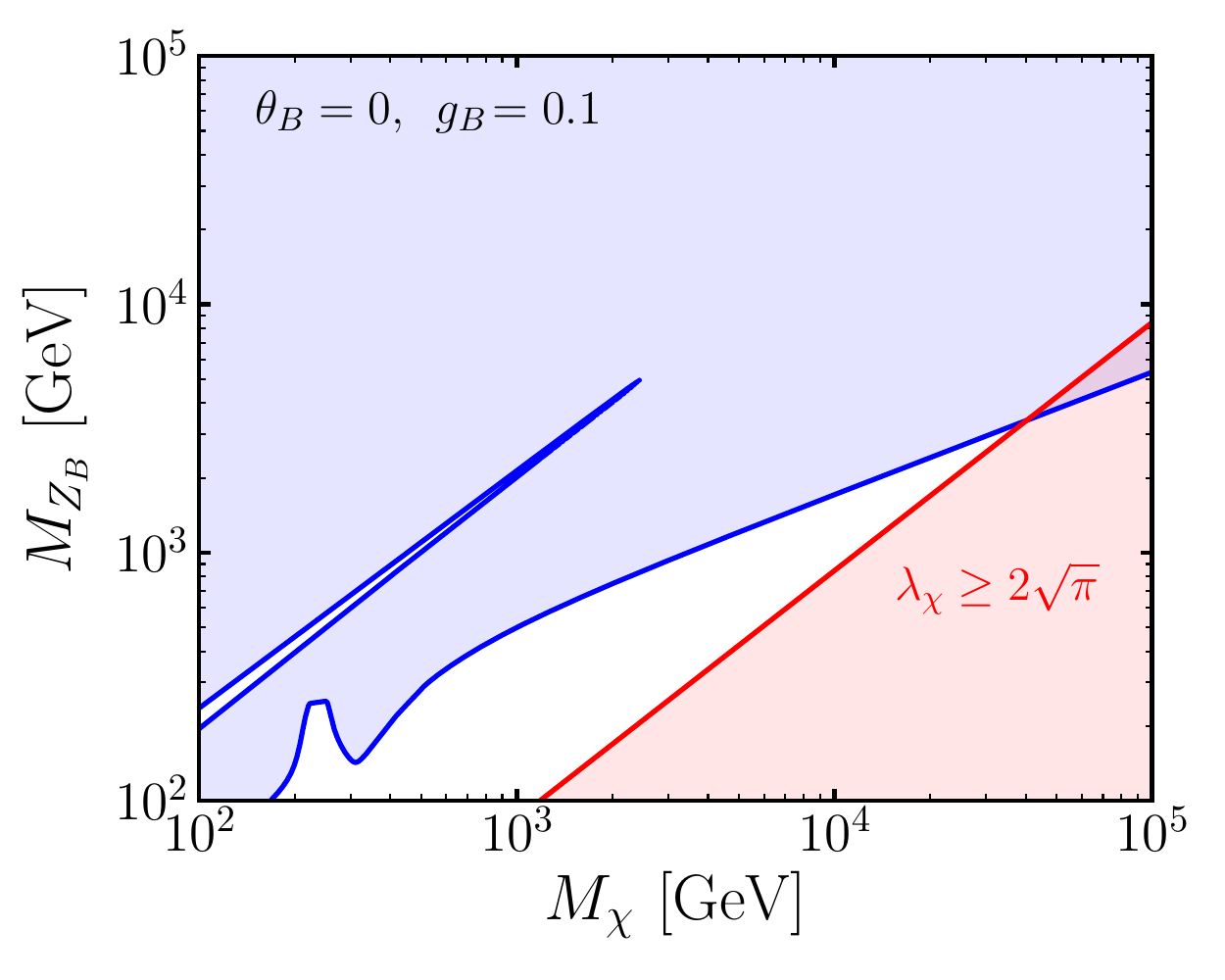} 
\includegraphics[width=0.48\linewidth]{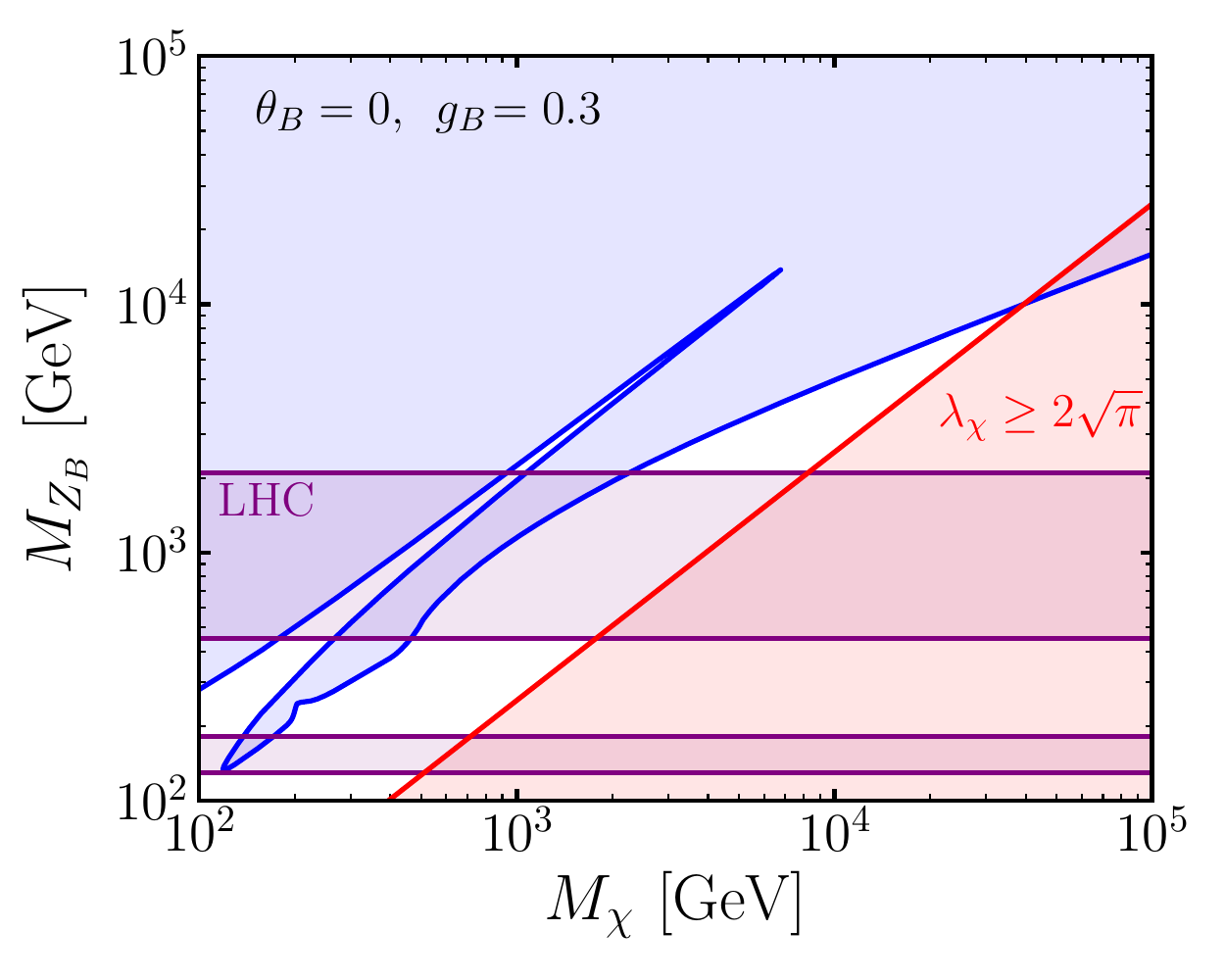} 
\includegraphics[width=0.48\linewidth]{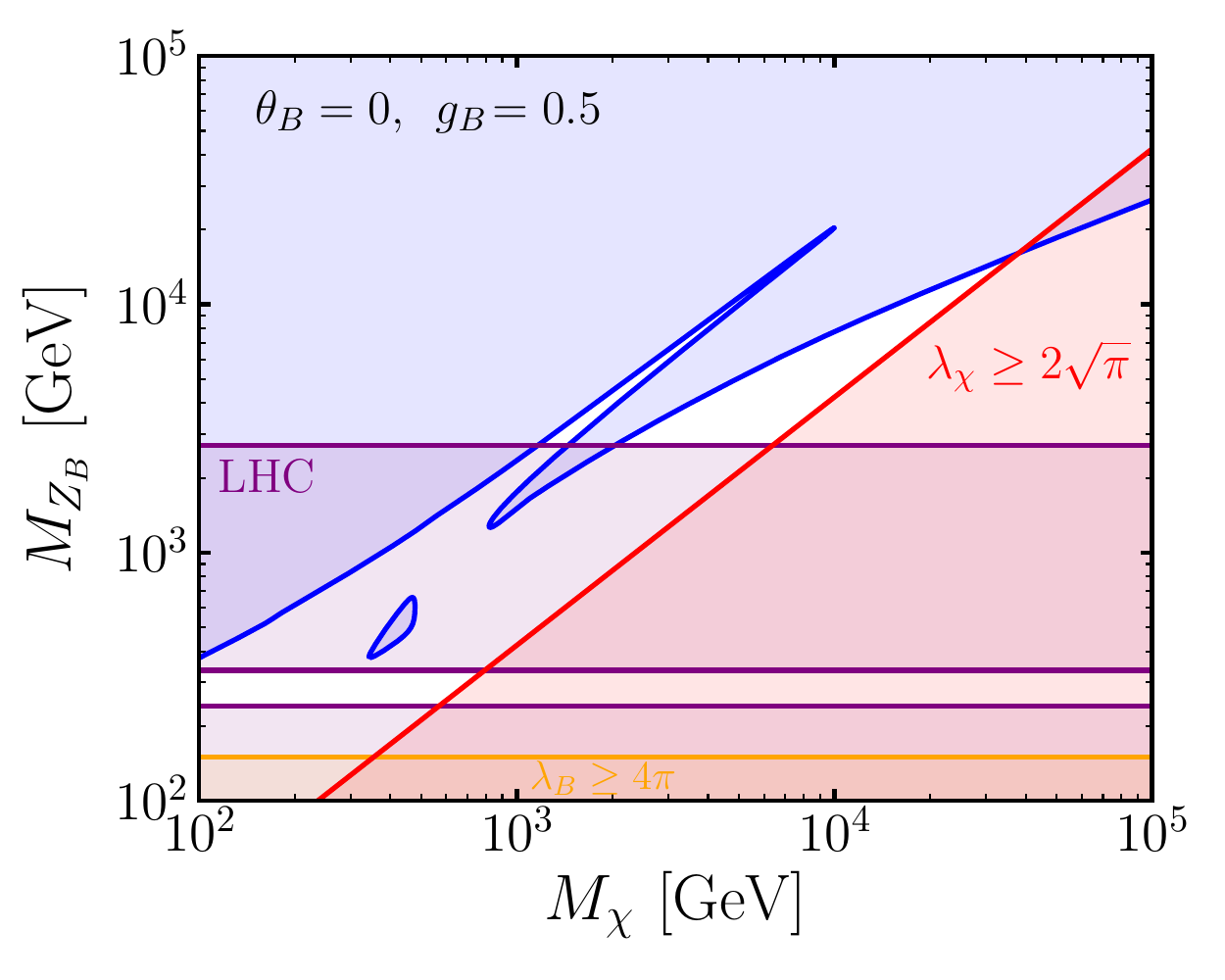} 
\includegraphics[width=0.48\linewidth]{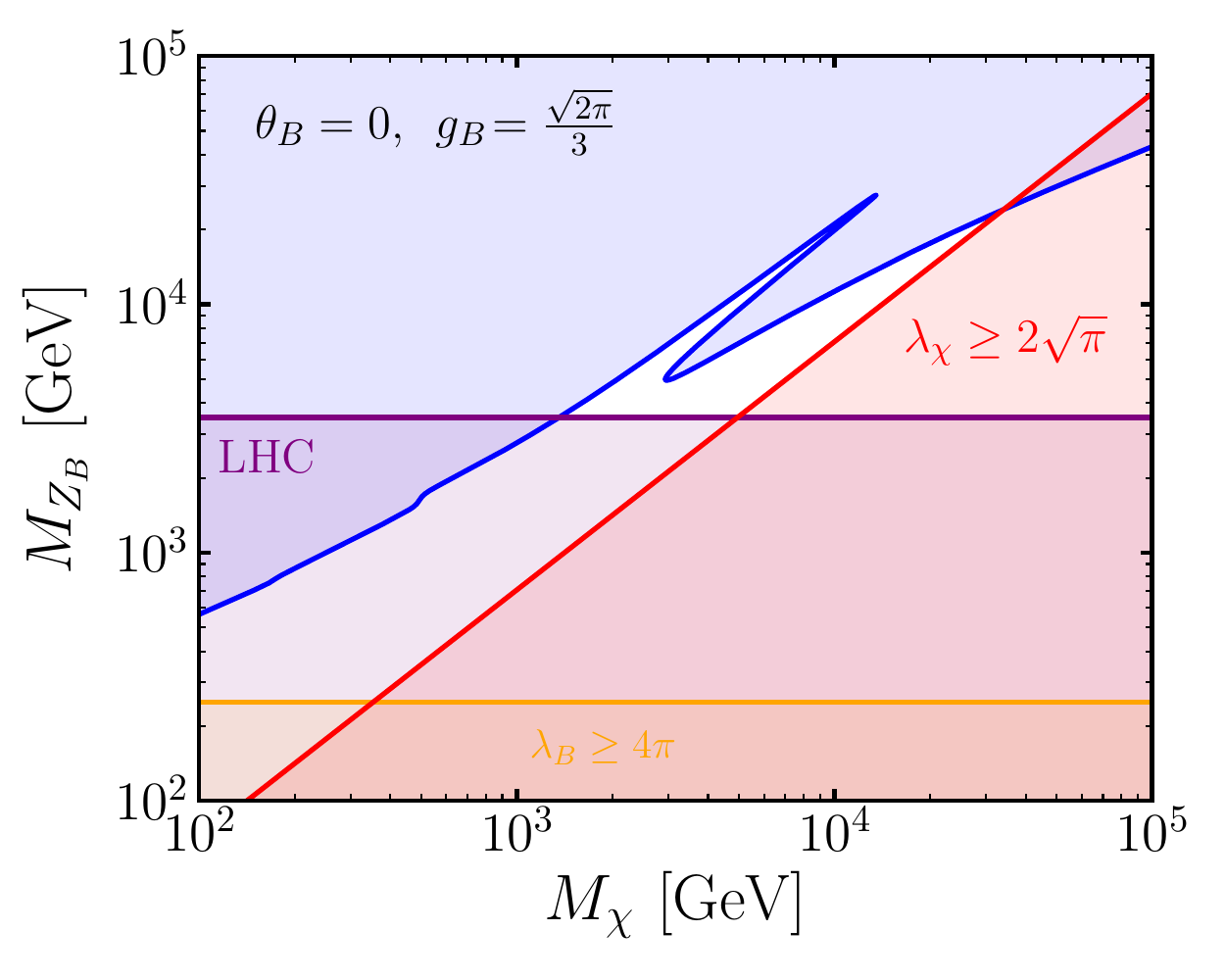} 
\caption{Parameter space allowed by the relic density constraint, LHC bounds and perturbative bounds for four different scenarios, $g_B=0.1$ (top-left), $g_B=0.3$ (top-right), $g_B=0.5$ (bottom-left) 
and $g_B=\sqrt{2 \pi}/3$ (bottom-right). We take $M_{h_2}=500$ GeV and no mixing angle. The region shaded in blue overproduces dark matter $ \Omega_{\rm{DM}} h^2 > 0.12$ and the region in red (yellow) is excluded by the perturbative bound on the Yukawa coupling $\lambda_\chi$ (scalar coupling $\lambda_B$). 
The horizontal purple bands are excluded by the LHC bounds on the leptophobic gauge boson mass.}
\label{Relic-scenarios}
\end{figure}

In Fig.~\ref{Relic-scenarios} we present results for the relic density in the $M_\chi$ versus $M_{Z_{B}}$ plane. The dark blue line corresponds to the measured relic abundance by Planck satellite measurement of $\Omega_{\rm{DM}} h^2 = 0.1197 \pm 0.0022$ \cite{Ade:2015xua}. The region shaded in light blue overproduces the dark matter relic density and it is ruled out unless the thermal history of the Universe is modified.
We show results for four different scenarios, $g_B=0.1$ (top-left), $g_B=0.3$ (top-right), $g_B=0.5$ (bottom-left) and $g_B=\sqrt{2 \pi}/3$ (bottom-right). 

\begin{figure}[t]
\centering
\includegraphics[width=0.48\linewidth]{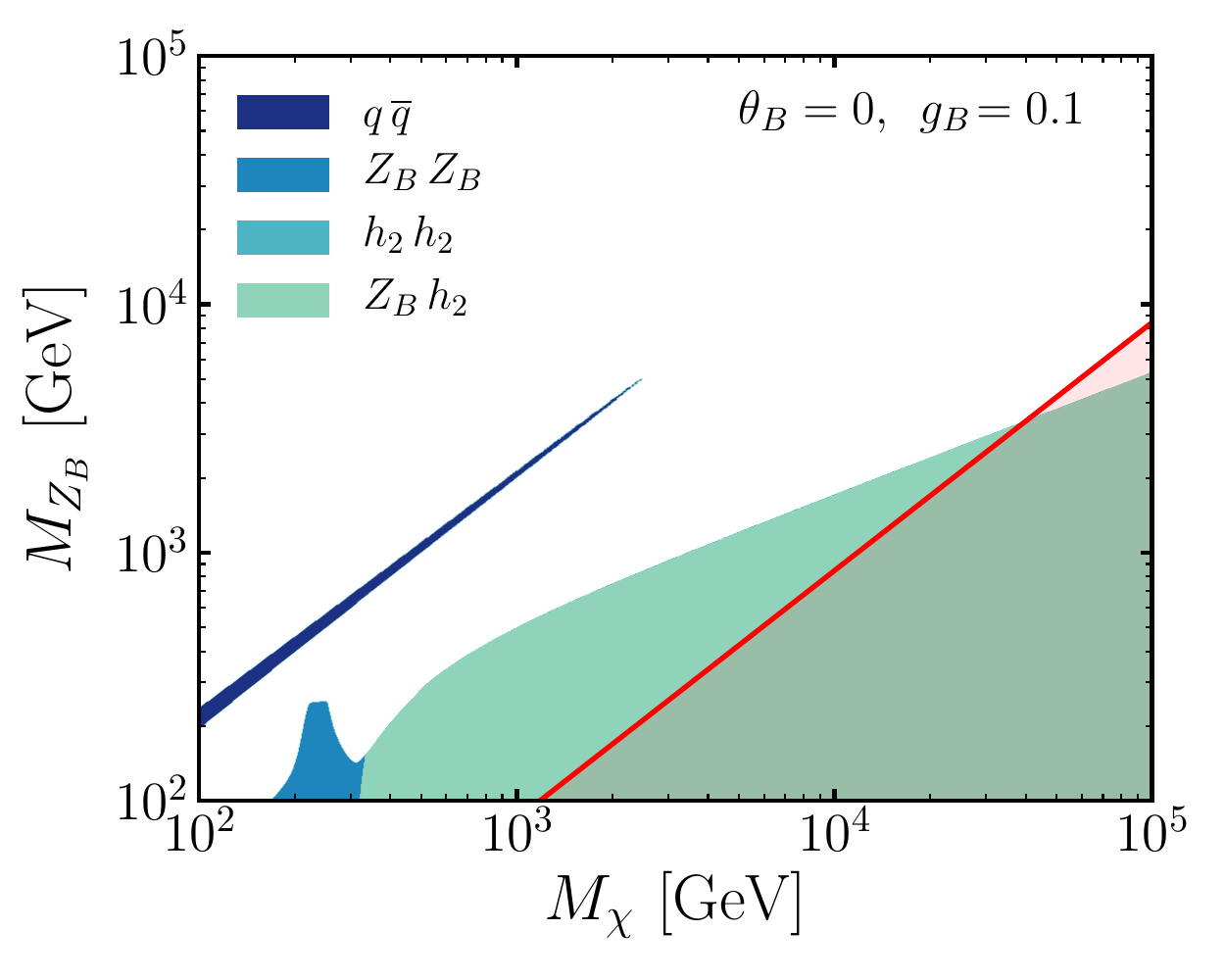} 
\includegraphics[width=0.48\linewidth]{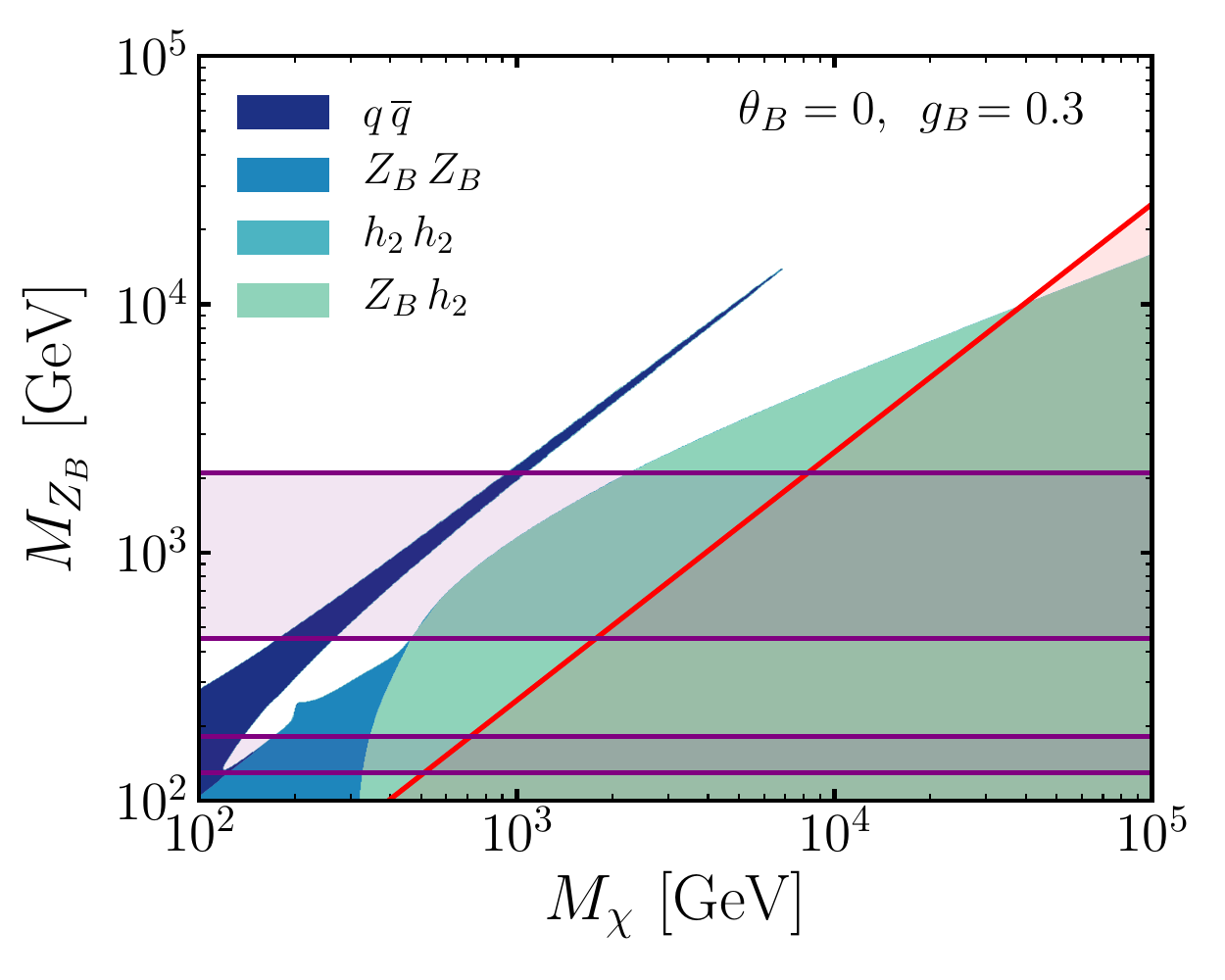} 
\includegraphics[width=0.48\linewidth]{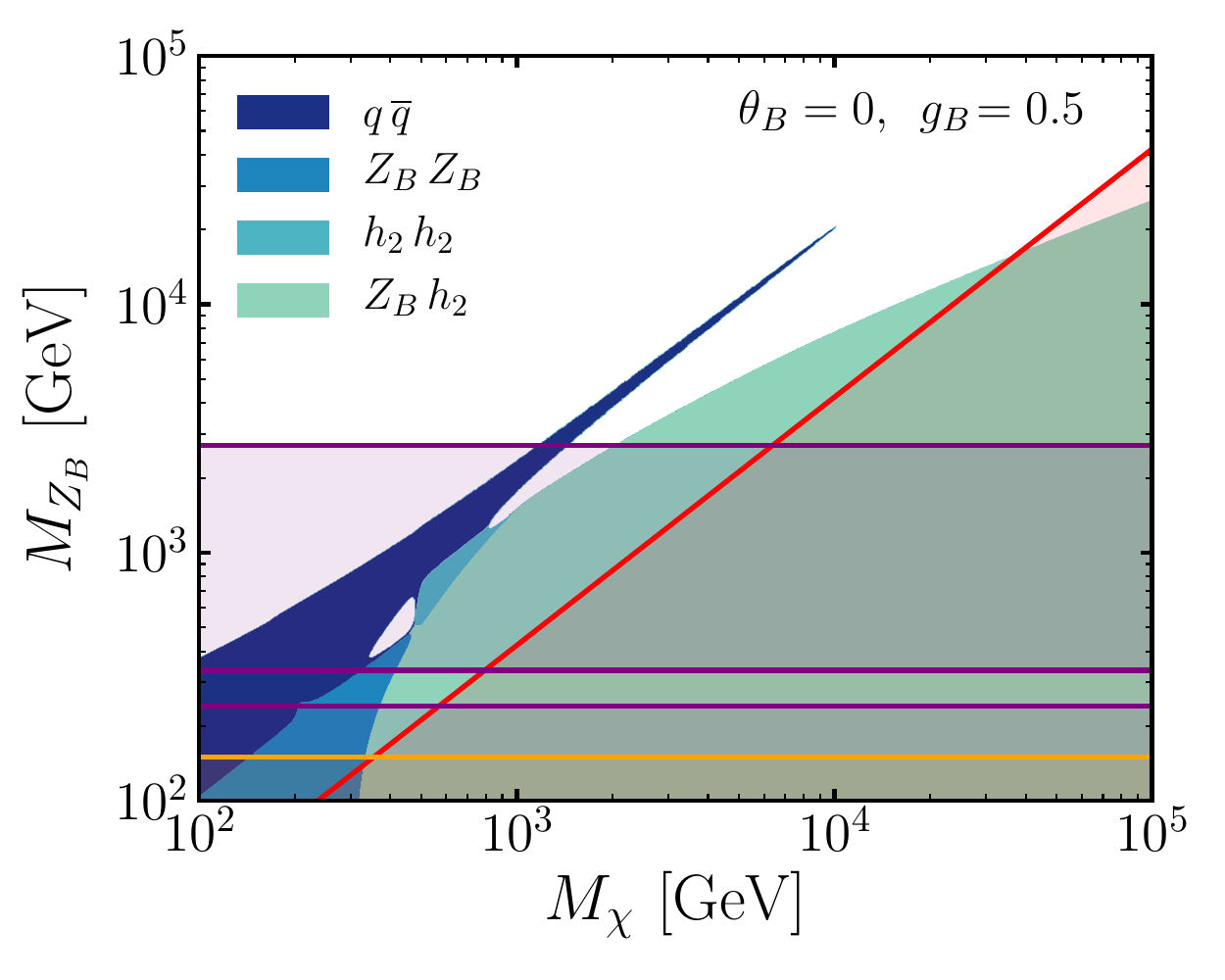} 
\includegraphics[width=0.48\linewidth]{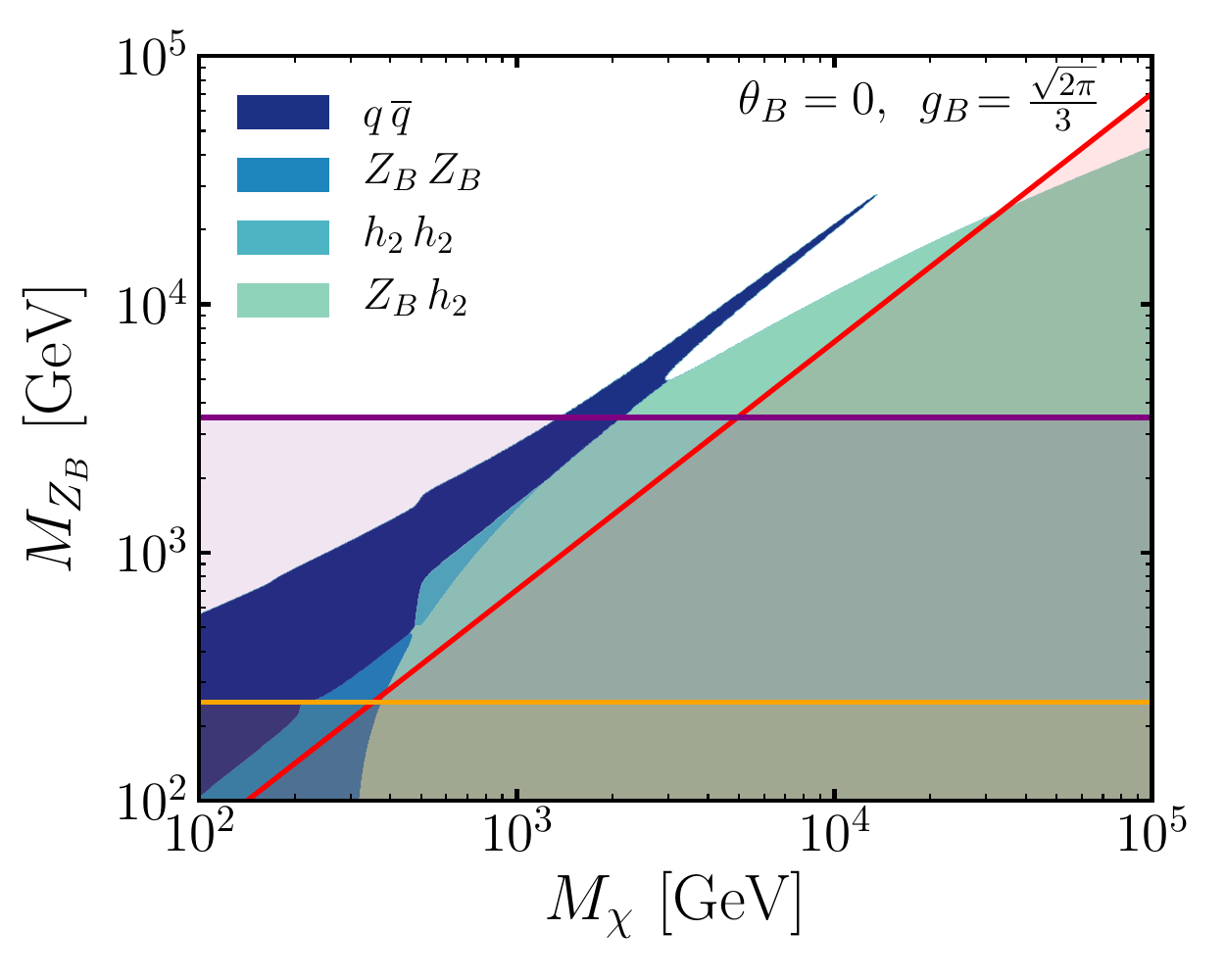} 
\caption{Regions in the $M_\chi$ versus $M_{Z_{B}}$ plane that satisfy the relic density constraint $\Omega_{\rm{DM}} h^2 \leq 0.12$, the colors indicate which annihilation channel is the dominant one. We fix $M_{h_2}=500$ GeV and $\theta_B=0$. The horizontal purple bands are excluded by dijet searches at the LHC. The region in red (yellow) is excluded by the perturbative bound on the Yukawa coupling $\lambda_\chi$ (scalar quartic coupling $\lambda_B$).}
\label{Relic-channels}
\end{figure}

The $Z_B$ mediator has direct coupling to quarks and would appear as a resonance in dijet searches at the LHC. In our work, we apply the bounds from CMS and ATLAS~\cite{FileviezPerez:2018jmr} and present the excluded regions in purple bands in Figs.~\ref{Relic-scenarios} and \ref{Relic-channels}. These bounds have a strong dependence on the coupling $g_B$ and disappear for $g_B \leq 0.1$. Additionally, the leptophobic gauge boson will develop a kinetic mixing with hypercharge (and hence the SM $Z$ boson) from radiative corrections. This, in turn, will modify electroweak precision observables which have been measured to great accuracy. However, as has been shown in \cite{Ellis:2018xal}, these bounds become relevant only for $M_{Z_B} \lesssim 100$ GeV; therefore, we do not consider them for our analysis. The mass of the second Higgs is fixed to $M_{h_2}=500$ GeV and the mixing angle to $\theta_B=0$. The region shaded in red is ruled out by the perturbative bound on the Yukawa coupling, $\lambda_\chi < 2\sqrt{\pi}$.  Similarly, the yellow region is ruled out by the perturbative bound on the scalar quartic coupling, $\lambda_B < 4\pi$. 

%We will further discuss this in Section \ref{sec:UpperBound}. 
The choice of the perturbative limit $g_B=\sqrt{2 \pi}/3$ is responsible for the upper bound on the symmetry breaking scale $M_{Z_B} \leq 28$ TeV. This striking feature tells us that one can hope to test or rule out this theory in the near future. The choice of zero scalar mixing can be motivated as follows. In the SM, the electroweak phase transition occurs around $T_{\rm EW} \approx 160$ GeV. For masses above the TeV scale (which is the region preferred after considering LHC bounds), we find the dark matter freeze-out temperature to be $x_f\approx 26-29$. Consequently, for $M_\chi \gtrsim 5$ TeV the freeze-out temperature is above the electroweak phase transition, and hence, at the time when DM freezes out the Higgs field has zero vacuum expectation value and there is no scalar mixing. It is important to mention that the bounds from the relic density constraint are very similar in the case where one considers a non-zero mixing angle.

In order to develop a better understanding of our results, we present in Fig.~\ref{Relic-channels} the parameter space allowed by the relic density constraint. We color each region depending on which annihilation is the dominant one; namely, the one that gives the largest contribution to the relic abundance at freeze-out. The region in dark blue corresponds to the region in parameter space where annihilation into quarks is the dominant channel, this happens near the resonance $M_\chi \approx M_{Z_B}/2$. 
The region in light blue is where the annihilation into $Z_B Z_B$ is the most dominant one. This occurs due to the resonance  $\chi \chi \rightarrow h_2^* \rightarrow Z_B Z_B$ when $M_\chi \approx M_{h_2}/2 \approx 250$ GeV. This resonant behavior can be easily appreciated in the upper left panels in Figs.~\ref{Relic-scenarios} and~\ref{Relic-channels}. The resonance is cut close to the diagonal $M_\chi \approx M_{Z_B}$, because the $Z_B Z_B$ channel becomes kinematically closed above.

%In the dark green region the $h_2 \, h_2$ channel is the dominant one; however, this is the case only in a small region in the parameter space. 
The region in which the $h_2 \, h_2$ annihilation channel is the dominant one is colored in dark green. The latter is velocity-suppressed and at freeze-out $v_{\rm DM} \approx 0.3$; consequently, this channel is dominant only in a small region in the parameter space. Once the $Z_B h_2$ channel becomes kinematically open, it becomes the dominant channel, as illustrated by the region in light green. 
%Moreover, this channel is responsible for the upper bound on the dark matter mass. 
As can be appreciated the upper bound on the gauge boson mass, $M_{Z_B} \leq 28$ TeV, is defined by the annihilation into quarks, while the upper bound on the dark matter 
mass, $M_{\chi} \leq 34$ TeV, is determined by the $Z_B h_2$ channel. 

%%%%%%%%%%%%%%%%%%%%%
\subsection{Direct Detection}
%%%%%%%%%%%%%%%%%%%%% 
Our dark matter candidate can interact with quarks in the nucleon via exchange of a leptophobic gauge boson or one of the physical scalars through the Higgs mixing. These two processes do not interfere with each other and the spin-independent cross-section between $\chi$ and a nucleon can be written as $\sigma_{\chi N}^\text{TOT} =  \sigma_{\chi N}(Z_B) + \sigma_{\chi N}(h_i)$, where
\begin{eqnarray}
\sigma_{\chi N}^\text{SI}(Z_B)&=& \frac{27}{8 \pi}\frac{g_B^4 M_N^2}{M_{Z_B}^4} v^2,\\
\sigma_{\chi N}^\text{SI}(h_i)&=&\frac{72 G_F}{\sqrt{2} 4 \pi}\sin^2 \theta_B \cos^2 \theta_B M_N^4 \frac{g_B^2 M_\chi^2}{M_{Z_B}^2}\left(\frac{1}{M_{h_1}^2}-\frac{1}{M_{h_2}^2}\right)^2 f_N^2,
\end{eqnarray}
$M_N$ corresponds to the nucleon mass, $G_F$ is the Fermi constant  and for the effective Higgs-nucleon-nucleon coupling we take $f_N=0.3$ \cite{Hoferichter:2017olk}. The axial coupling between $\chi$ and $Z_B$ leads to velocity suppression of the cross-section and we can write,
\begin{equation}
\sigma_{\chi N}^\text{TOT} = \sigma_{\chi N}(h_i) + \sigma_{\chi N}^\text{0}(Z_B) v^2.
\end{equation}
The bounds coming from direct detection experiments are obtained under the assumption that the leading order in the cross-section is velocity independent. In order to apply these bounds we proceed as follows,
\begin{equation}
 \sigma_{\chi N}(h_i) + \sigma^0_{\chi N}(Z_B) v_\text{eff}^2 \leq \sigma_{\chi N}^{\text{DDexp}},
\end{equation}
where  $\sigma_{\chi N}^{\text{DDexp}}$ is the upper bound on the scattering cross-section given by the direct detection experiments, and the effective velocity is given by the ratio $\overline{v^3}/\overline{v}$, where the average velocity is the velocity of the dark matter convoluted with a Maxwell-Boltzmann distribution. We find that $v_\text{eff} \approx 0.001 \, c $.
\begin{figure}[tbp]
\centering
\includegraphics[width=0.99\linewidth]{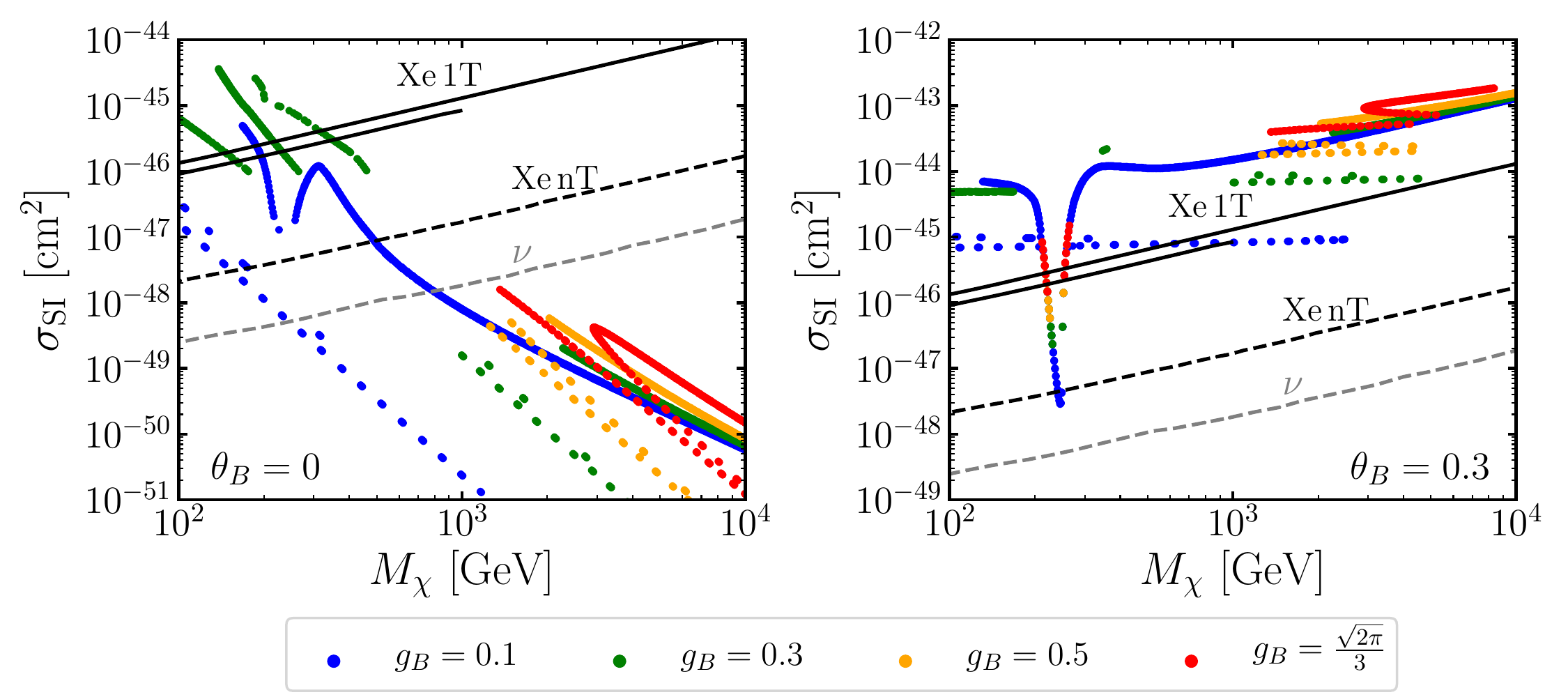} 
\caption{Predictions for the direct-detection spin-independent cross-section as a function of the dark matter mass. In the left (right) panel we present the predictions for $\theta_B=0$ ($\theta_B=0.3$). All points agree with the measured relic abundance by the Planck satellite $\Omega_{\rm{DM}}  h^2 = 0.1197 \pm 0.0022$ \cite{Ade:2015xua} and satisfy constraints from the LHC. The solid black lines show current experimental bounds from Xenon-1T \cite{Aprile:2017iyp, Aprile:2018dbl}, the dashed black line shows the projected sensitivity for Xenon-nT \cite{Aprile:2015uzo} and the dashed gray line shows the coherent neutrino scattering limit \cite{Billard:2013qya}. }
\label{fig:DD}
\end{figure}

Collider searches of a new scalar that mixes with the Higgs combined with measurements of Higgs properties provide constraints on the mixing angle \cite{Duerr:2017whl, Ilnicka:2018def}. In our study, the mass of the second Higgs is above the electroweak scale, and hence, we take the bound $\sin \theta_B \leq 0.3$. In Fig.~\ref{fig:DD} we present the predictions for the spin-independent cross-section for different values of the gauge coupling. In order to select the points we proceed as follows. First, we select those points that give the measured relic abundance of $\Omega_{\rm{DM}} h^2 = 0.1197 \pm 0.0022$. Then, we remove those points that do not satisfy LHC and/or perturbativity bounds. This is the reason behind the discontinuities in the points shown.

The left panel shows the results for $\theta_B=0$ and, as expected, due to the velocity suppression the points with dark matter mass above the TeV scale lie below the neutrino floor and will escape detection from future experiments. In contrast, for the maximal  mixing allowed, $\theta_B=0.3$, all the points that saturate the relic density are ruled out by direct detection bounds except for those that lie close to the $h_2$ resonance ($M_\chi \approx M_{h_2}/2$), this is shown in the right panel in Fig.~\ref{fig:DD}. For intermediate values of the mixing angle, the predictions lie in between these two regions, and thus, some of these points will be probed by future direct detection experiments, such as Xenon-nT \cite{Aprile:2015uzo}.

%%%%%%%%%%%%%%%%%%%%%%%%%%%
\section{GAMMA LINES FROM LEPTOPHOBIC DM}
\label{sec:GammaLines}
%%%%%%%%%%%%%%%%%%%%%%%%%%%%  
The model discussed so far suffers from gauge anomalies. In this section, we will argue that the extra fermions required to make the theory anomaly-free also lead to dark matter annihilation into photons. 
Consequently, the photon spectrum will contain features that could be observed in future telescopes. 
In order to predict the cross-sections for
%for the dark matter annihilation into gamma rays for the channels,
$$\chi \chi \to \gamma \gamma, \, \gamma Z, \, \gamma h, $$
we need to study anomaly-free gauge theories based on $\U(1)_B$. These processes are quantum mechanical and can be predicted only after we understand the anomaly cancellation. In this section, we study the predictions for the gamma lines in the simple models proposed in Refs.~\cite{Duerr:2013dza,Perez:2014qfa}.

Gamma lines are typically suppressed with respect to other processes contributing to the continuum, such as final state radiation (FSR), and hence, it is hard to observe them. However, it is possible to have scenarios in which the processes that contribute to the continuum close to the gamma lines are highly suppressed and the gamma lines become visible. This can be easily understood in terms of the energy of the gamma lines, 
%
%Typically the gamma lines are suppressed with respect other processes contributing to the continuum 
%spectrum and it is impossible to observe those. However, one can have scenarios where the processes contributing to the continuum close to the gamma 
%lines are highly suppressed and then the gamma lines can be visible. There is a simple way to understand this issue if we think of the energy of the gamma lines
%
\begin{equation}
E_i^\gamma = M_{\rm{DM}} \left( 1 - \frac{M_i^2}{4 M_{\rm{DM}}^2}\right),
\end{equation}
where $M_i=0,M_Z,M_h$ for the corresponding gamma lines, ${\rm{DM}} \  {\rm{DM}} \to \gamma \gamma, \gamma Z, \gamma h$.
At the same time, there are processes such as final state radiation processes,  ${\rm{DM}} \  {\rm{DM}} \to \rm{SM} \ \rm{SM} \ \gamma$, which occur at tree level 
and could spoil the visibility of the gamma lines because the maximal energy of the photon in this case is given by
\begin{equation}
E_{\rm max}^\gamma = M_{\rm{DM}} \left( 1 - \frac{M_{\rm{SM}}^2}{M_{\rm{DM}}^2}\right).
\end{equation}
Thus, when $M_{\rm{DM}} \gg M_{\rm{SM}}$ the gamma line visibility is spoiled if the FSR processes are not suppressed because $E_{\rm max}^\gamma \approx E_i^\gamma$.
There are many studies in the literature where the predictions for the gamma spectrum from dark matter annihilation have been investigated; we refer the reader to Ref.~\cite{Bringmann:2012ez} for a detailed discussion of different scenarios. Recently, we have investigated dark matter models where 
the gamma lines are visible due to the fact that the FSR processes are naturally suppressed~\cite{Duerr:2015vna,FileviezPerez:2019rcj}. 
For studies of models that give rise to gamma-ray lines see Refs.~\cite{Jackson:2009kg,Dudas:2012pb,Jackson:2013pjq,Jackson:2013rqp,Dudas:2013sia,Hooper:2014fda}.

In this article, we investigate the predictions for gamma lines in simple extensions of the Standard Model, our DM candidate is a Majorana fermion and the FSR processes are velocity-suppressed. 
Consequently, there is hope to observe the gamma lines, which is a striking feature of these models. There is a simple way to understand the velocity suppression of some DM annihilation channels.
In our case the DM annihilation can be mediated only by the $Z^{'}$ and the Higgses. For example, for the ${\rm{DM}}  \  {\rm{DM}} \to \bar{q} q$ channel the relevant operator is 
$(\bar{\chi} \gamma^\mu \gamma^5 \chi) (\bar{q} \gamma_\mu q)$ and the amplitude is proportional to $\bar{v}_\chi(p_2) \gamma^\mu \gamma^5 u_\chi(p_1)$. In the non-relativistic 
limit, the term $\bar{v}_\chi(p_2) \gamma^i \gamma^5 u_\chi(p_1)$ is proportional to the velocity $v^i$, while $\bar{v}_\chi(p_2) \gamma^0 \gamma^5 u_\chi(p_1)$ is not velocity suppressed.
However, since $\bar{u}_q (p_4) \gamma_0 v_q(p_3)=0$ when $\vec{p}_3=-\vec{p}_4$ the full amplitude is velocity suppressed.  In general one can say that the amplitude of any process, 
${\rm{DM}}  \  {\rm{DM}} \to (Z^{'})^* \to \rm{any}$, is velocity suppressed if there is no contribution from the term $\bar{v}_\chi(p_2) \gamma^0 \gamma^5 u_\chi(p_1)$. 

For the gamma lines, 
${\rm{DM}}  \  {\rm{DM}} \to (Z^{'})^* \to \gamma X$, where $X=\gamma, h, Z$, the above argument can be applied. These processes are not velocity 
suppressed when the amplitude is proportional to $\bar{v}_\chi(p_2) \gamma_0 \gamma^5 u_\chi(p_1) \delta \Gamma^{0 \sigma \rho}_{Z^{'} X \gamma}$, 
where $\delta \Gamma^{\mu \sigma \rho}_{Z^{'} X \gamma}$ is the effective $Z^{'} X \gamma$ coupling computed in Ref.~\cite{Duerr:2015wfa}. Moreover, this effective vertex violates parity, and hence, it can only be obtained from fermions in the loop that have an axial interaction with the new gauge boson, i.e. $\bar{\psi}\gamma^\mu\gamma^5\psi Z'_\mu$. This simple argument 
can be applied to any annihilation channel with any number of fields in the final state. See also Ref.~\cite{Kumar:2013iva} for the 
study of properties of the amplitude in different  DM models. 
%

%%%%%%%%%%%%%%%%%%%%%%
\subsection{Theories for Baryon Number}
%%%%%%%%%%%%%%%%%%%%%%
We have discussed the main features of simplified models with a baryonic Majorana dark matter. This type of dark matter candidate 
has been predicted in models where the baryon number is defined as a local symmetry. The simplest models based on $\U(1)_B$ with a Majorana dark matter have been pointed out in Refs.~\cite{Duerr:2013dza,Perez:2014qfa}. In Tables~\ref{tabla1} and ~\ref{tabla2}, we show the extra fermionic representations in these theories.
\begin{itemize}
\item {{Model I}}: Ref.~\cite{Duerr:2013dza} proposed a simple anomaly-free theory for the spontaneous breaking of local baryon number where the anomalies 
are canceled with the fields listed in Table~\ref{tabla1}.

\begin{table}[h]\setlength{\bigstrutjot}{6pt}
\centering
\begin{tabular}{|ccccc|}\hline
    Fields & $\SU(3)_C$ & $\SU(2)_L$ & $\U(1)_Y$  & $\U(1)_B$ \bigstrut\\\hline\hline
    $\Psi_L = \mqty(\Psi_L^0 \\ \Psi_L^-)$ & ${1}$ & ${2}$ & $-\frac{1}{2}$ & $-\frac{3}{2}$\bigstrut\\
    $\Psi_R = \mqty(\Psi_R^0 \\ \Psi_R^-)$ & $1$ & $2$ & $-\frac{1}{2}$ & $\frac{3}{2}$  \bigstrut\\
    $\eta_R^-$ & $ 1$ & $ 1$ & $-1$ & $-\frac{3}{2}$ \bigstrut\\
    $\eta_L^-$ & $1$ & $1$ & $-1$ & $\frac{3}{2}$ \bigstrut\\
    $\chi_R^0$ & $1$ & $1$ & $0$ & $-\frac{3}{2}$ \bigstrut\\
    $\chi_L^0$ & $1$ & $1$ & $0$ & $\frac{3}{2}$ \bigstrut\\
\hline
  \end{tabular}
  \caption{Fermionic representations in the model proposed in Ref.~\cite{Duerr:2013dza}.}
  \label{tabla1}
\end{table}
The relevant Lagrangian for our discussion is given by
\begin{eqnarray}
{\cal{L}} &\supset& y_1 \bar{\Psi}_L H \eta_R + y_2 \bar{\Psi}_L \tilde{H} \chi_R + y_3 \bar{\Psi}_R H \eta_L + y_4 \bar{\Psi}_R \tilde{H} \chi_L  \nonumber \\
&+& \lambda_1 \bar{\Psi}_L \Psi_R S_B^* + \lambda_2 \bar{\eta}_R \eta_L S_B^* + \lambda_3 \bar{\chi}_R \chi_L S_B^* + \frac{\lambda_\chi}{\sqrt{2}}  \chi_L \chi_L S_B^* + \lambda_5 \chi_R \chi_R S_B + \rm{h.c.},
\label{LagrangianModelI}
\end{eqnarray}
where $H \sim (1,2,1/2, 0)$ and $S_B \sim (1,1,0,3)$. In general, the DM candidate can be a linear combination of all neutral fields 
and it is a Majorana fermion. Recently, this model has been investigated in Ref.~\cite{Caron:2018yzp}. See also a recent 
study in Ref.~\cite{FileviezPerez:2018jmr} where one can have a similar model where the DM candidate is a Dirac fermion.
\vspace{0.5cm}
\item {{Model II}}: In the model proposed in Ref.~\cite{Perez:2014qfa} the anomalies are canceled with only four extra fields listed in Table~\ref{tabla2}.
\begin{table}[h]\setlength{\bigstrutjot}{6pt}
\centering
\begin{tabular}{|ccccc|}\hline
    Fields & $\SU(3)_C$ & $\SU(2)_L$ & $\U(1)_Y$  & $\U(1)_B$ \bigstrut\\\hline\hline
    $ \Psi_L = \mqty(\Psi_L^+ \\ \Psi_L^0)$ & ${1}$ & ${2}$ & $\frac{1}{2}$ & $\frac{3}{2}$\bigstrut\\
    $ \Psi_R = \mqty(\Psi_R^+ \\ \Psi_R^0)$ & $1$ & $2$ & $\frac{1}{2}$ & $-\frac{3}{2}$  \bigstrut\\
    $\Sigma_L = \frac{1}{\sqrt{2}} \mqty(\Sigma^0_L & \sqrt{2}\Sigma^+_L \\ \sqrt{2}\Sigma^-_L & -\Sigma^0_L)$ & $1$ & $3$ & $0$ & $-\frac{3}{2}$ \bigstrut\\
   $\chi_L^0$ & $1$ & $1$ & $0$ & $-\frac{3}{2}$ \bigstrut\\ \hline
%     $S_B$ & $\tb 1$ & $\tb 1$ & $0$ & $3$ 
%\bigstrut\\ \hline
  \end{tabular}
  \caption{Fermionic representations in the model proposed in Ref.~\cite{Perez:2014qfa}.}
  \label{tabla2}
\end{table}%

The Yukawa interactions needed to generate vector-like masses for the new fermions are given by 
\begin{eqnarray}
{\cal{L}} &\supset& y_1 \bar{\Psi}_R H \chi_L + y_2  H^\dagger \Psi_L \chi_L + y_3 H^\dagger \Sigma_L \Psi_L + y_4 \bar{\Psi}_R \Sigma_L  H \nonumber \\
&+& \lambda_1 \bar{\Psi}_L \Psi_R S_B + \frac{\lambda_\chi}{\sqrt{2}} \chi_L \chi_L S_B + \lambda_\Sigma \text{Tr} \Sigma_L^2 S_B + \rm{h.c.},
\label{LagrangianModelII}
\end{eqnarray}
the DM is always a Majorana fermion and it can be a linear combination of all neutral fields present in the theory. 
See Ref.~\cite{Ohmer:2015lxa} for a detailed phenomenological study of this model.
\end{itemize}
In these theories, the stability of the dark matter candidate is a natural consequence of the symmetry breaking.
In this context the local $\U(1)_B$ is broken to an $Z_2$ discrete symmetry which protects the stability of the DM candidate.
It is important to mention that the stability of the DM candidate is never spoiled by any higher-dimensional operators one could write.

%%%%%%%%%%%%%%%%%%%%%%
\subsection{Predictions for Gamma Lines}
%%%%%%%%%%%%%%%%%%%%%%
%
\begin{figure}[tb]
\centering
\includegraphics[width=0.3\linewidth]{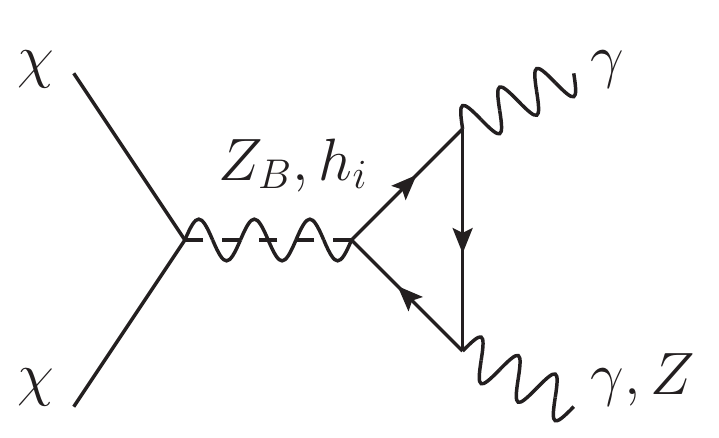} 
\includegraphics[width=0.3\linewidth]{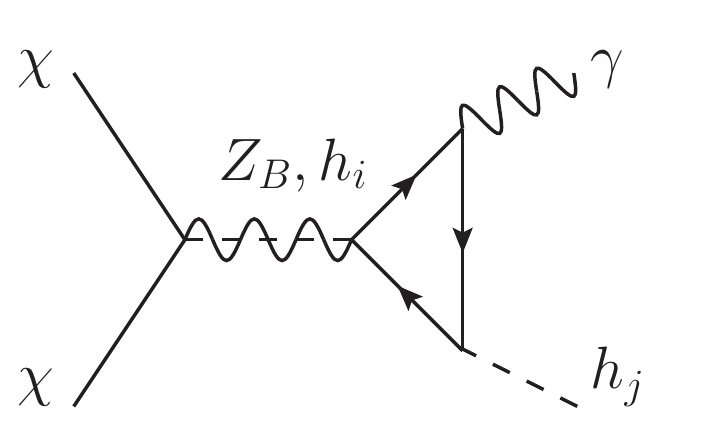}  \\
\includegraphics[width=0.3\linewidth]{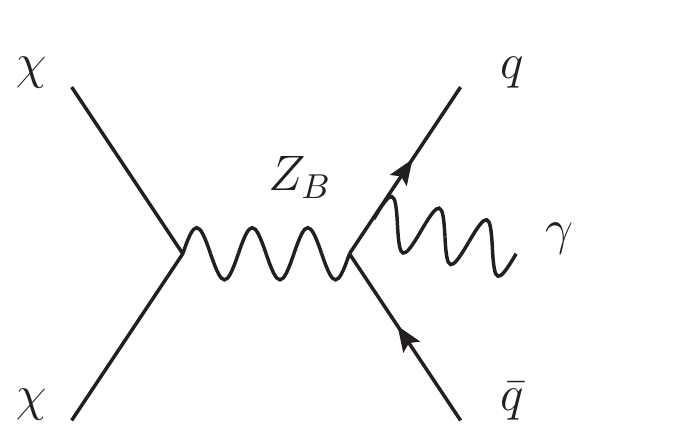} 
\includegraphics[width=0.3\linewidth]{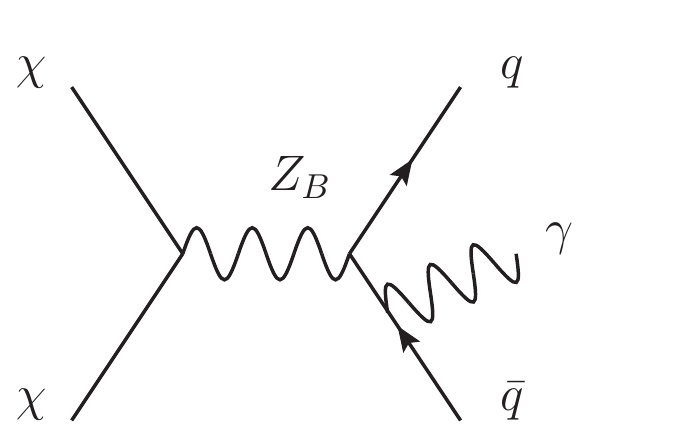} 
\caption{Upper panel: Feynman graphs for the dark matter annihilation into gamma rays. Lower panel: Feynman graphs for the processes that contribute to final state radiation.}
\label{GammaGraph}
\end{figure}
The relevant interactions for the DM annihilation into gamma lines can be written as
\begin{equation}
{\cal L} \supset -g_B \overline{f_+}\left(n_V^{f_+} \gamma^\mu + n_A^{f_+} \gamma^\mu \gamma^5\right) f_+ Z^B_\mu - \frac{e}{\sin \theta_W \cos \theta_W} \overline{f_+} \left(g_V^f \gamma^\mu + g_A^f \gamma^\mu \gamma_5\right) f_+ Z_\mu,
\end{equation}
where $f_+$ are the charged fermions entering in the loops shown in the upper panel of Fig.~\ref{GammaGraph}. This figure also shows the three different possibilities for the generation of gamma lines, $\chi \chi \to \gamma \gamma$, $\chi \chi \to Z \gamma$ and $\chi \chi \to h_i \gamma$. 
%
%%%%%%%%%%%%%%%%%%
\subsubsection{Gamma Lines}
%%%%%%%%%%%%%%%%%%
In these theories, the DM annihilation into a photon and a Higgs is velocity-suppressed but the cross-section for the gamma lines coming from $\chi \chi \to \gamma \gamma$ and $\chi \chi \to \gamma Z$ can be large. These cross-sections are given by
\begin{equation}
 \sigma v ({{\chi}\chi \to \gamma \gamma}) =\frac{\alpha^2}{\pi^3} \frac{g_B^4 n_\chi^2 M_\chi^2}{M_{Z_B}^4}\frac{(4M_\chi^2 - M_{Z_B}^2)^2}{(4M_\chi^2-M_{Z_B}^2)^2+\Gamma_{Z_B}^2M_{Z_B}^2} \left | \sum_{f_+} N_c^{f_+} n_A^{f _+}Q_{f_+}^2M_{f_+}^2 C_0^\gamma \right |^2,
\label{ACgg}
\end{equation}
and
\begin{equation}
\sigma v (\chi \chi \to \gamma Z) =\frac{\alpha^2 \, g_B^4 n_\chi^2 }{32 \pi^3 \sin^2 2\theta_W}\frac{(4M_\chi^2-M_Z^2)^3}{ (4M_\chi^2-M_{Z_B}^2)^2+\Gamma_{Z_B}^2 M_{Z_B}^2}\frac{(M_{Z_B}^2-4M_\chi^2)^2}{M_\chi^4M_{Z_B}^4} \left| \sum_f N_c^{f_+} Q_{f_+}  n_A^{f_+} g_{V}^{f_+} 2M_{f_+}^2 C_0^Z \right|^2,
\label{ACgZ}
\end{equation}
where $\alpha=e^2/4\pi$, $C_0^A = C_0(0,M_A^2,s; M_f,M_f,M_f)$ is the Passarino-Veltman loop function as defined in Ref.~\cite{Patel:2015tea} and $n_\chi$ is the dark matter charge under $\U(1)_B$. In order to predict the cross-sections for these quantum mechanical processes we need to use the interactions 
of the new fields required for anomaly cancellation.
\vspace{0.2cm}
\begin{itemize}
\item {Model I}: In this case, as the following diagrams show,
%%%%%%%%%%%%%%%%%%%%%%%%%%%%%%%%%%%%%%%%%%%%%%%%%%%%%
\newline
\begin{center}
\includegraphics[scale=1.0]{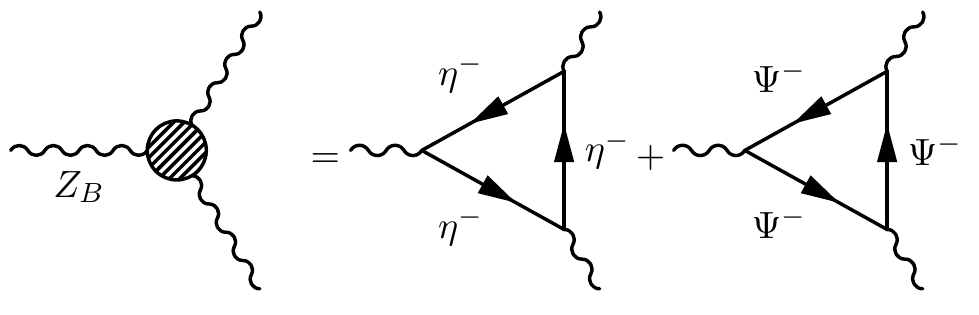}
\end{center}
%%%%%%%%%%%%%%%%%%%%%%%%%%%%%%%%%%%%%%%%%%%%%%%%%%%%%
only the charged fields $\Psi^-$ and $\eta^-$ will contribute to the DM annihilation into gamma lines. The relevant Lagrangian in this case is given by
\begin{eqnarray}
{\cal L}^{\rm I} & \supset&  e  \, \overline{\Psi^-} \slashed{A} \Psi^- + e \, \overline{\eta^-}\slashed{A}\eta^- + \frac{e}{\tan 2 \theta_W} \overline{\Psi^-}\slashed{Z}\Psi^-  \nonumber \\
&& - e \tan \theta_W \overline{\eta^-}\slashed{Z} \eta^- -\frac{3}{2}g_B \overline{\Psi^-}\slashed{Z}_B \gamma_5 \Psi^- +\frac{3}{2} g_B \overline{\eta^-}\slashed{Z}_B \gamma_5 \eta^-,
\end{eqnarray}
from which one can read the couplings to the $Z_B$ gauge boson: $n_V^{\Psi^-} = 0$, $n_A^{\Psi^-}=3/2$, $n_V^{\eta^-}=0$ and $n_A^{\eta^-}=-3/2$, and the couplings to the $Z$ boson: $g_V^{\Psi^-}=-\frac{1}{2}$, $g_A^{\Psi^-}=0$, $\displaystyle g_V^{\eta^-}= \sin^2 \theta_W$ and $g_A^{\eta^-}=0$. 
In this scenario, in the limit where the couplings $y_1=y_2=y_3=y_4=0$ the masses for the new charged fermions read as 
\begin{equation}
M_{\Psi^-} = \frac{\lambda_1 M_{Z_B}}{3\sqrt{2} g_B}, \quad M_{\eta^-} = \frac{\lambda_2 M_{Z_B}}{3\sqrt{2}g_B}.
\end{equation}
The upper bound on $M_{Z_B}$ derived from relic density constraints defines a global upper bound for the new charged fermion masses since the Yukawa couplings $\lambda_1$ and $\lambda_2$ are bounded by perturbativity. Moreover, the perturbative bounds define an upper bound for each mass of the new mediator. Hence, for a given $M_{Z_B}$ and $M_\chi$, the masses of the charged fermions are constrained to be  in the following range,
\begin{equation}
M_\chi < M_{f^+} \leq \frac{2 \sqrt{\pi}}{3g_B}M_{Z_B},
\label{ConPer}
\end{equation}
where $f^+$ symbolizes any of the charged new fermions; the lower bound comes from ensuring the stability of the dark matter candidate whereas the upper bound comes from perturbativity on the Yukawa couplings. We note that the dark matter candidate and the charged fermions satisfy the same perturbative bound.

In Fig.~\ref{gL_M1} we present our predictions for the DM annihilation into gamma lines for different choices of $g_B$ in the context of Model I for the two relevant annihilation channels. All the points shown saturate the relic abundance, $\Omega_{\rm{DM}} h^2 = 0.1197 \pm 0.0022$, and satisfy LHC and direct detection constraints.
From Eqs.~\eqref{ACgg} and \eqref{ACgZ} we can see that when the charged fermions have the same mass, the cross-section vanishes. We have taken the maximum mass splitting by choosing $M_{\Psi^-} = 1.3 M_\chi$, in order to ensure the stability of the dark matter candidate, and $M_{\eta^-} = 2 \sqrt{\pi}M_{Z_B}/3g_B$, i.e. the largest value allowed by perturbativity. Additionally, for each point we check that the perturbativity constraint, Eq.~\eqref{ConPer}, is satisfied.

\begin{figure}[t]
\centering
\includegraphics[width=0.99\linewidth]{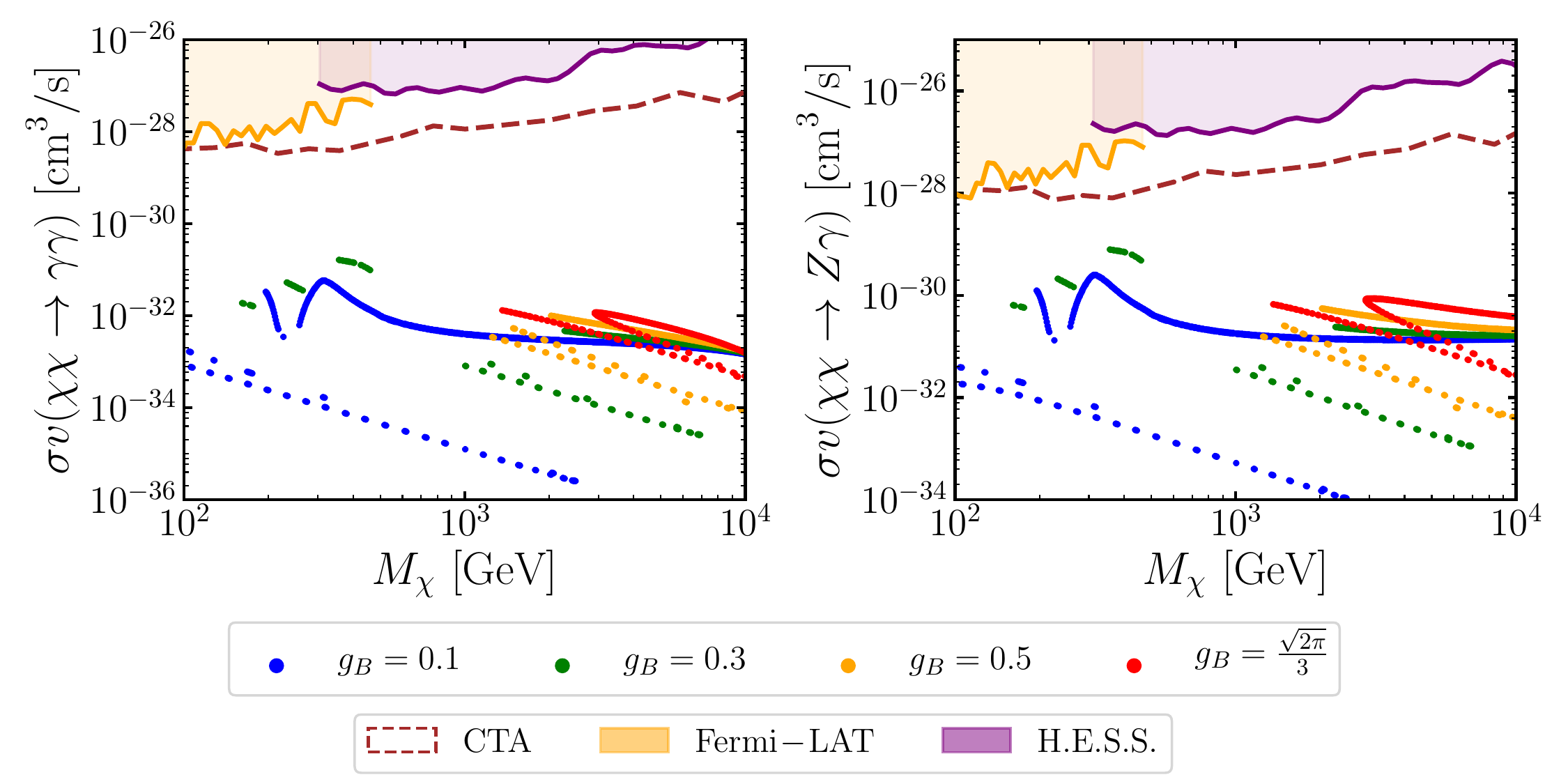} 
\caption{Predictions for the DM annihilation into two photons (left panel) and a photon and a $Z$ boson (right panel) in the context of Model I. We set $M_{h_2}=500$ GeV and $\theta_B=0$. The value of $M_{Z_B}$ is chosen such that every point satisfies $\Omega_{\rm{DM}} h^2 = 0.1197 \pm 0.0022$, different colors correspond to different values of the gauge coupling $g_B$. All points shown pass LHC and direct detection constraints. The regions colored in yellow are purple show the excluded parameter space by the Fermi-LAT \cite{Ackermann:2015lka,Ackermann:2013uma} and H.E.S.S. \cite{Abramowski:2013ax}, respectively. The dashed brown line shows the projected sensitivity from the CTA collaboration \cite{Acharya:2017ttl}.}
\label{gL_M1}
\end{figure}

\item {Model II}: In this context, the relevant Lagrangian for the gamma lines reads as
\begin{eqnarray}
{\cal L}^{\rm II} & \supset&  -e \overline{\Sigma^+}\slashed{A}\Sigma^+ - e\overline{\Psi^+}\slashed{A} \Psi^+-\frac{e}{\tan \theta_W}\overline{\Sigma^+}\slashed{Z}\Sigma^+ - \frac{e}{\tan 2 \theta_W}\overline{\Psi^+}\slashed{Z}\Psi^+ \nonumber \\
&& + \frac{3}{2}g_B\overline{\Psi^+}\slashed{Z}_B\gamma_5 \Psi^+ - \frac{3}{2}g_B \overline{\Sigma^+}\slashed{Z}_B \gamma_5 \Sigma^+,
\end{eqnarray}
where $\Sigma^+$ and $\Psi^+$ correspond to the fields contributing to DM annihilation into gamma lines
%%%%%%%%%%%%%%%%%%%%%%%%%%%%%%%%%%%%%%%%%%%%%%%%%%%%%
\newline
\begin{center}
\includegraphics[scale=1.0]{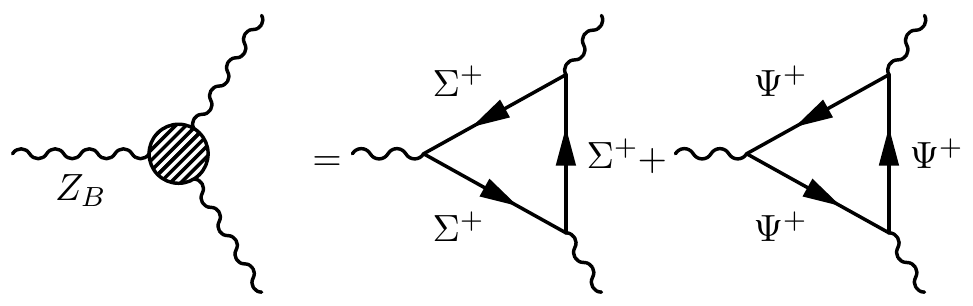}
\end{center}
%%%%%%%%%%%%%%%%%%%%%%%%%%%%%%%%%%%%%%%%%%%%%%%%%%%%%
%
%
\begin{figure}[t]
\centering
\includegraphics[width=0.99\linewidth]{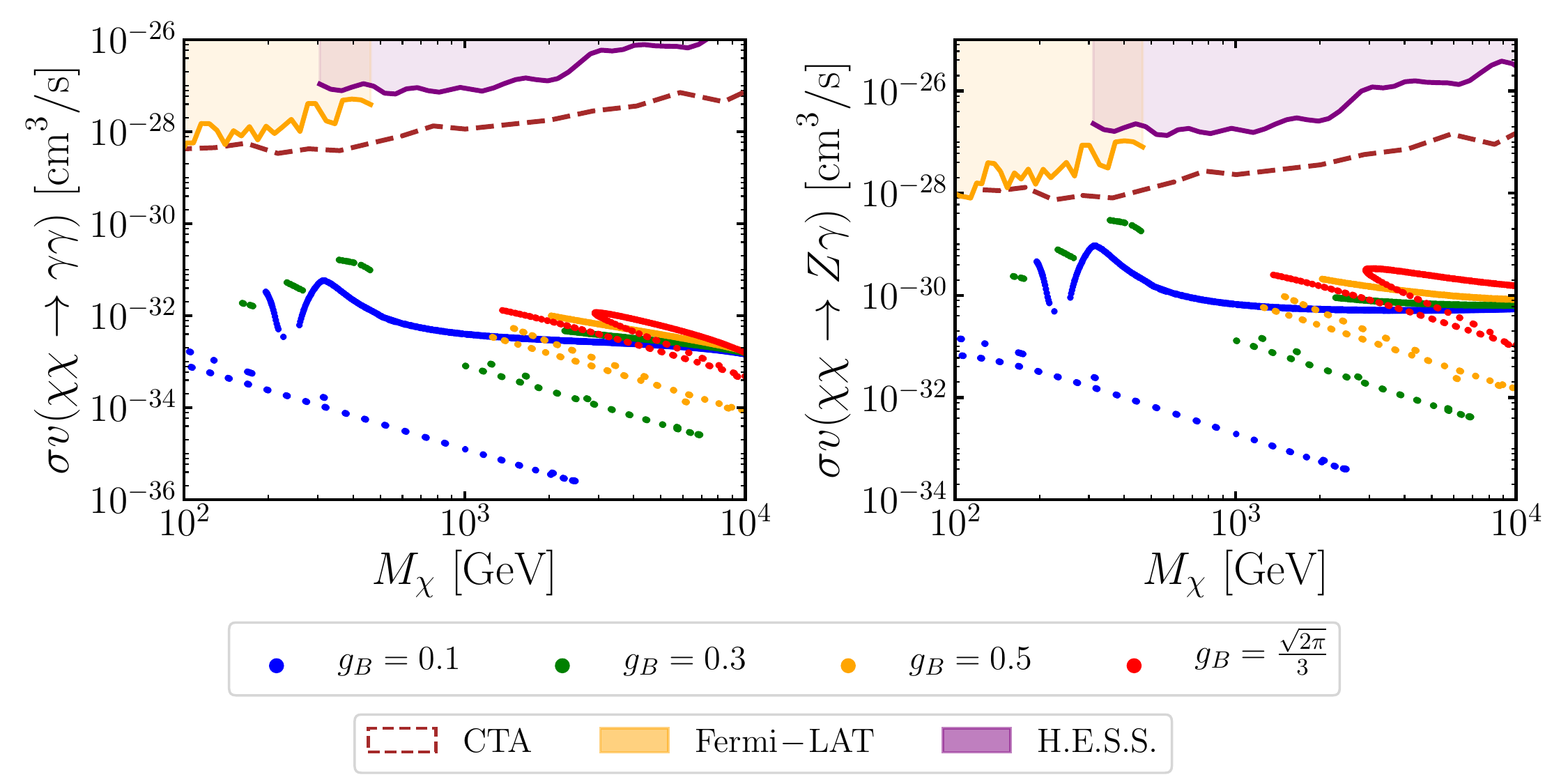} 
\caption{Predictions for the DM annihilation into two photons (left panel) and a photon and a $Z$ boson (right panel) in the context of Model II. We set $M_{h_2}=500$ GeV and $\theta_B=0$. The value of $M_{Z_B}$ is chosen such that every point satisfies $\Omega_{\rm{DM}} h^2 = 0.1197 \pm 0.0022$, different colors correspond to different values of the gauge coupling $g_B$. All points shown pass LHC and direct detection constraints. The regions colored in yellow are purple show the excluded parameter space by the Fermi-LAT \cite{Ackermann:2015lka,Ackermann:2013uma} and H.E.S.S. \cite{Abramowski:2013ax}, respectively. The dashed brown line shows the projected sensitivity from the CTA collaboration~\cite{Acharya:2017ttl}.}
\label{gL_M2}
\end{figure}
From the above Lagrangian, one can identify the couplings to the $Z_B$ boson: $n_V^{\Psi^+}=0$, $n_A^{\Psi^+}=-\frac{3}{2}$, $n_V^{\Sigma^+}=0$, $n_A^{\Sigma^+}=\frac{3}{2}$, and the couplings to the $Z$ boson: $g_V^{\Psi^+}=\frac{1}{2}$, $g_A^{\Psi^+}=0$, $g_V^{\Sigma^+}=\cos^2 \theta_W$ and $g_A^{\Sigma^+}=0$.
In this scenario, in the limit where the couplings $y_1=y_2=y_3=y_4=0$ the masses for the new charged fermions read as 
\begin{equation}
M_{\Psi^+} = \frac{\lambda_1 M_{Z_B}}{3\sqrt{2} g_B}, \quad M_{\Sigma^+} = \frac{\sqrt{2} \lambda_\Sigma M_{Z_B}}{3g_B}.
\end{equation}
As in the previous case, in here there is also a global upper bound on the charged fermion masses defined by the upper bound on the $M_{Z_B}$ and the perturbative bounds of the Yukawa couplings. As already discussed, the masses of the charged fermions must be in the range defined by Eq.~\eqref{ConPer} to ensure the stability of the dark matter candidate and the perturbativity of the Yukawa couplings $\lambda_\Sigma$ and $\lambda_1$. 

In Fig.~\ref{gL_M2} we present our predictions for the DM annihilation into gamma lines for different choices of $g_B$ in the context of Model II. As in Model I, the points satisfy $\Omega_{\rm{DM}} h^2 = 0.1197 \pm 0.0022$ and the masses of the charged fermions are chosen to maximize the splitting among them: $M_{\Psi^+} = 1.3 M_\chi$ and $M_{\Sigma^+} = 2\sqrt{\pi} M_{Z_B} /3 g_B$.
\end{itemize}

Regarding gamma lines, the only difference between both models is the coupling of one of the extra fields to the $Z$ boson. Consequently, for the same choice of masses, the $\gamma \gamma$ line is the same in the context of both models, but there are slight differences for the $\gamma Z$ line. 
The right panels in Figs.~\ref{gL_M1} and~\ref{gL_M2} show the cross-sections for the $Z \gamma$ line, as can be appreciated for Model I the predictions are slightly weaker than for Model II (due to the fact that $\eta^- \eta^- Z$ coupling is suppressed by $\tan^2\theta_W$ with respect to the $\Sigma^+ \Sigma^+ Z$ coupling, as shown in the above Lagrangians).
The left panels are the same because the $\gamma \gamma$ line dominates over the $Z \gamma$ line, so the predictions are the same for both models. 
The signals displayed in Figs.~\ref{gL_M1} and~\ref{gL_M2} are, unfortunately, in a region that remains a few orders of magnitude below from the current experimental sensitivity. However, we would like to remark that these two $\U(1)_B$ models are two of the most motivated SM extensions that predict axial interactions between charged fermions with the new mediator that allow the predictions for the DM annihilation into gamma lines, although being small, to be different from zero. Furthermore, there is hope that future telescopes can test or rule out these predictions. 
%
%%%%%%%%%%%%%%%%%%%%%%%%%%%%%%%%%%%%%%%
\begin{figure}[t]
\centering
\includegraphics[width=0.45\linewidth]{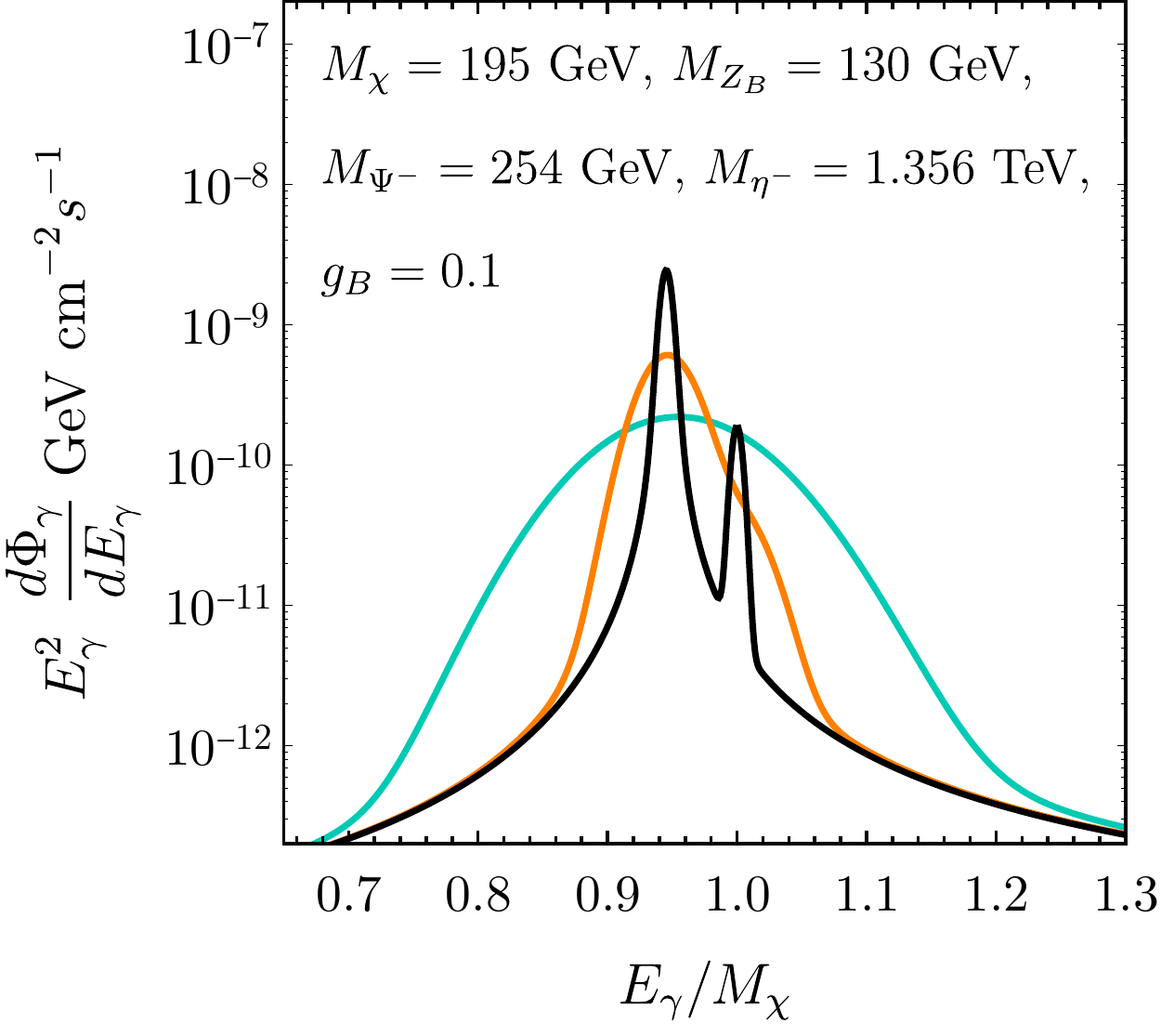} 
\includegraphics[width=0.45\linewidth]{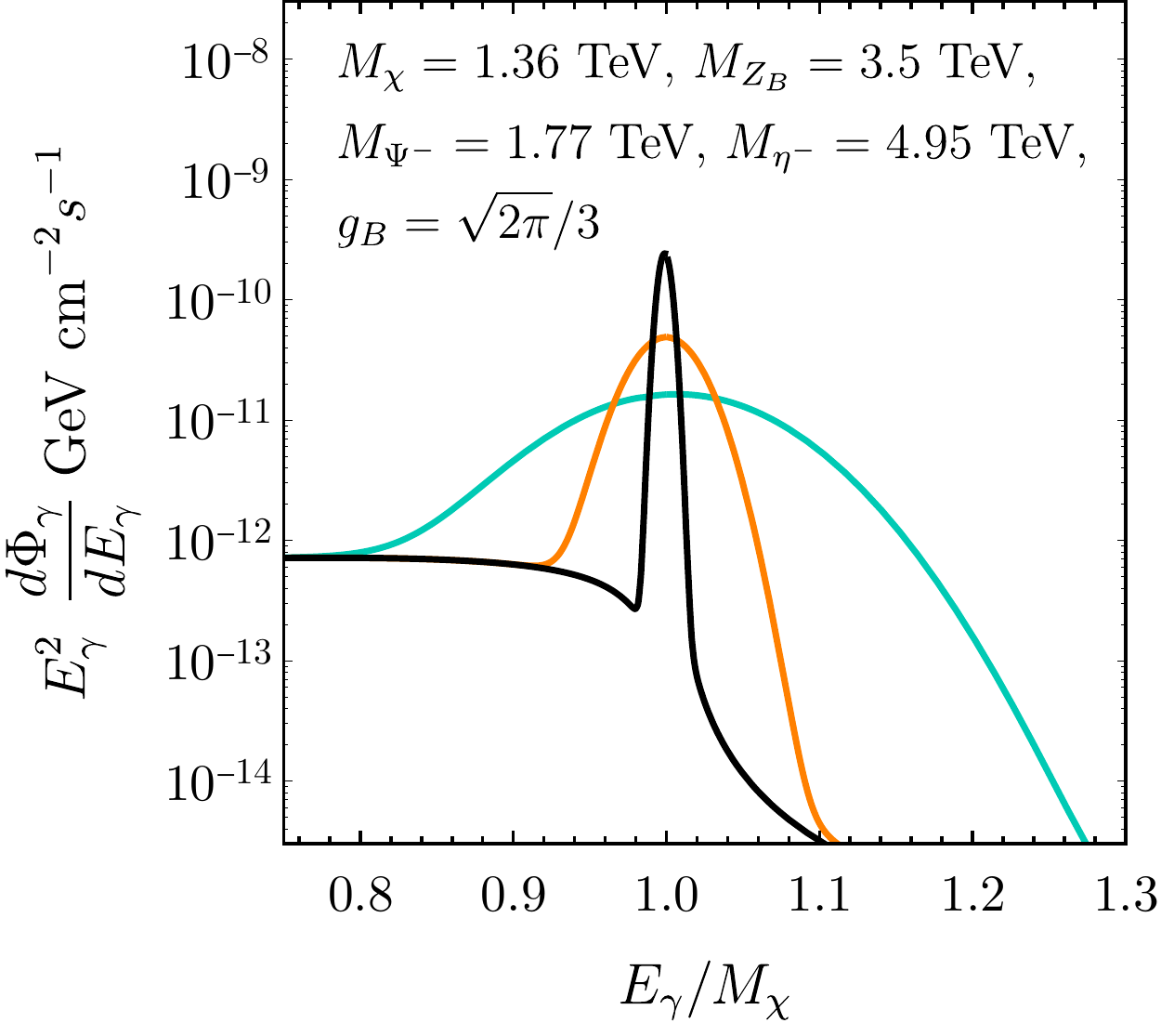} 
\hspace{1.3\textwidth} \includegraphics[width=0.4\linewidth]{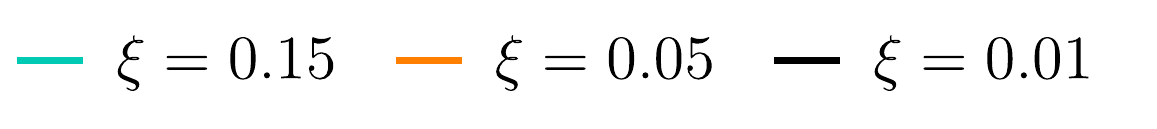} 
\caption{The differential spectrum of the dark matter annihilation into gamma rays for two different scenarios that satisfy the relic abundance measured by the Planck satellite $\Omega_{\rm{DM}} h^2 = 0.1197 \pm 0.0022$ \cite{Ade:2015xua} in Model I. We set $M_{h_2}=500$ GeV and $\theta_B=0$. Lines with different colors correspond to different energy resolutions $\xi = \{ 0.01, 0.05, 0.15\}$.}
\label{GLsignalsI}
\end{figure}
%%%%%%%%%%%%%%%%%%%%%%%%%%%%%%%%%%%%%%%%%
%%%%%%%%%%%%%%%%%%%%%%%%%%
\subsubsection{Final State Radiation Processes}
%%%%%%%%%%%%%%%%%%%%%%%%%%
As pointed out in Ref.~\cite{FileviezPerez:2019rcj}, a Majorana candidate for dark matter is crucial in order to see its annihilation into gamma lines. It is well known that the generation of gamma lines takes place through a quantum mechanical process whereas the processes contributing to the continuum spectrum are, in general, tree level processes (see lower panel of Fig.~\ref{GammaGraph}). However, if the DM candidate is a Majorana fermion, the amplitude for the processes contributing to the final state radiation can be expanded as 
\begin{equation}
|{\cal M}|^2_\text{FSR} = \frac{M_q^2}{M^2_{Z_B}}A +  v^2 B + {\cal O}(v^4),
\end{equation}
where A and B are coefficients defined at the end of this section. From here, one can see that they are either velocity or mass ($M_q / M_{Z_B}$) suppressed. Therefore, in this case, one can hope to distinguish the gamma lines from the continuum spectrum. The coefficients that parametrize the expansion in velocity for the FSR processes are given by
\begin{eqnarray}
&&A= 12 \pi \, \alpha \, g_B^4  Q_q^2   (M_{Z_B}^2-4M_\chi^2)^2 \frac{(E_q+E_\gamma - M_\chi)^2\left(2(E_q-M_\chi)(E_q + E_\gamma -M_\chi)-3M_q^2 \right) }{M_{Z_B}^2(E_q-M_\chi)^2(E_q+E_\gamma-M_\chi)^2((4M_\chi^2-M_{Z_B}^2)^2+\Gamma_{Z_B}^2M_{Z_B}^2)},  \\
&&B=12 \pi \, \alpha \,  g_B^4 M_\chi^2 Q_q^2  \times \nonumber \\
&&  \frac{ \left(2E_q M_\chi (E_\gamma^2-3 E_\gamma M_\chi + 2 M_\chi^2)-2 E_q^4 - 2 E_q^3 ( E_\gamma - 2 M_\chi) -E_q^2(E_\gamma^2-6E_\gamma M_\chi + 6M_\chi^2)-2M_\chi^2(E_\gamma - M_\chi)^2\right)}
{M_{Z_B}^2 (E_q+E_\gamma-M_\chi)^2((4M_\chi^2-M_{Z_B}^2)^2+\Gamma_{Z_B}^2M_{Z_B}^2)}.\nonumber \\
\end{eqnarray}
%
%%%%%%%%%%%%%%%%%%%%%%%%%%%%%%%%%%%%%%%%
\begin{figure}[t]
\centering
\includegraphics[width=0.45\linewidth]{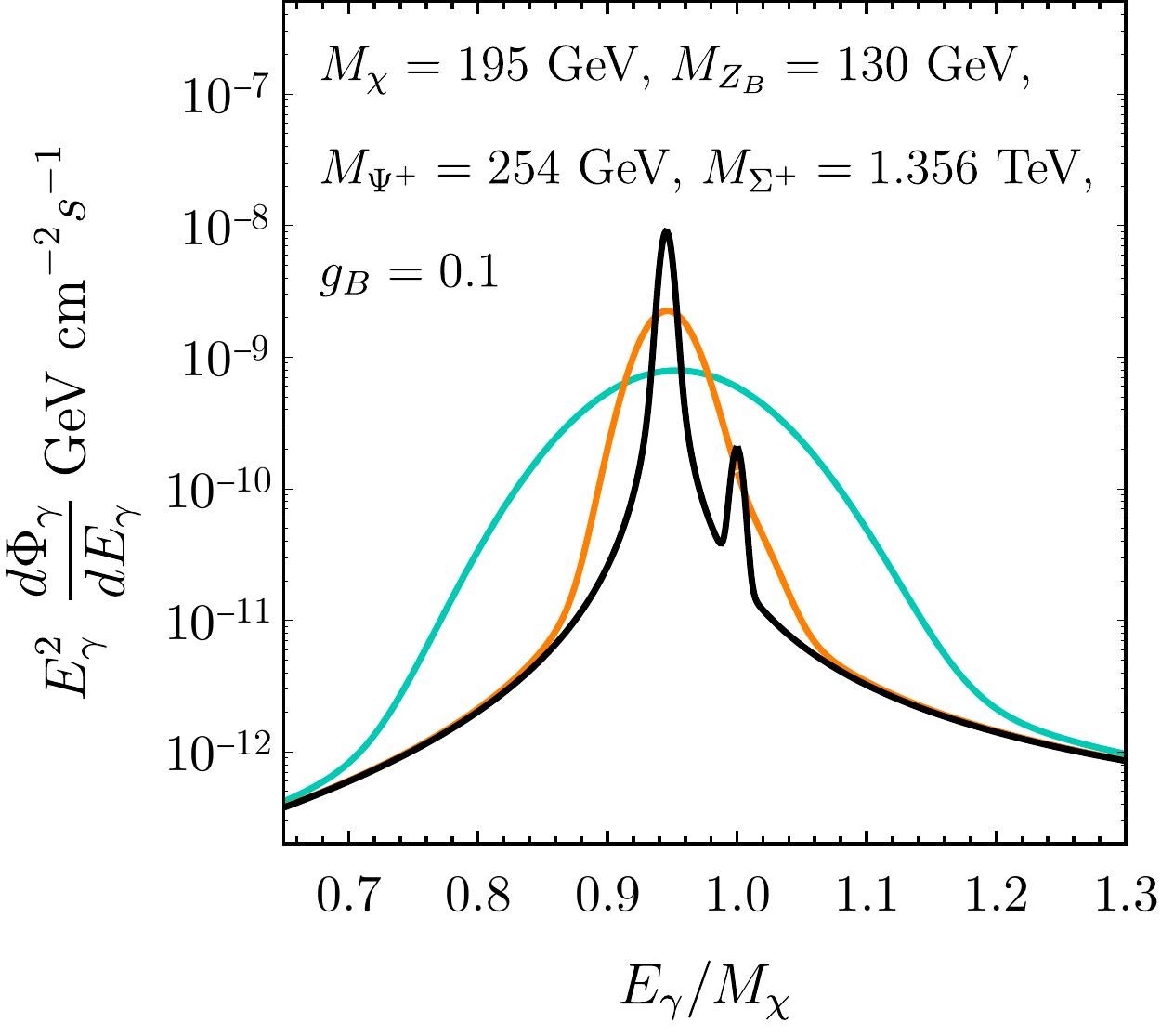} 
\includegraphics[width=0.45\linewidth]{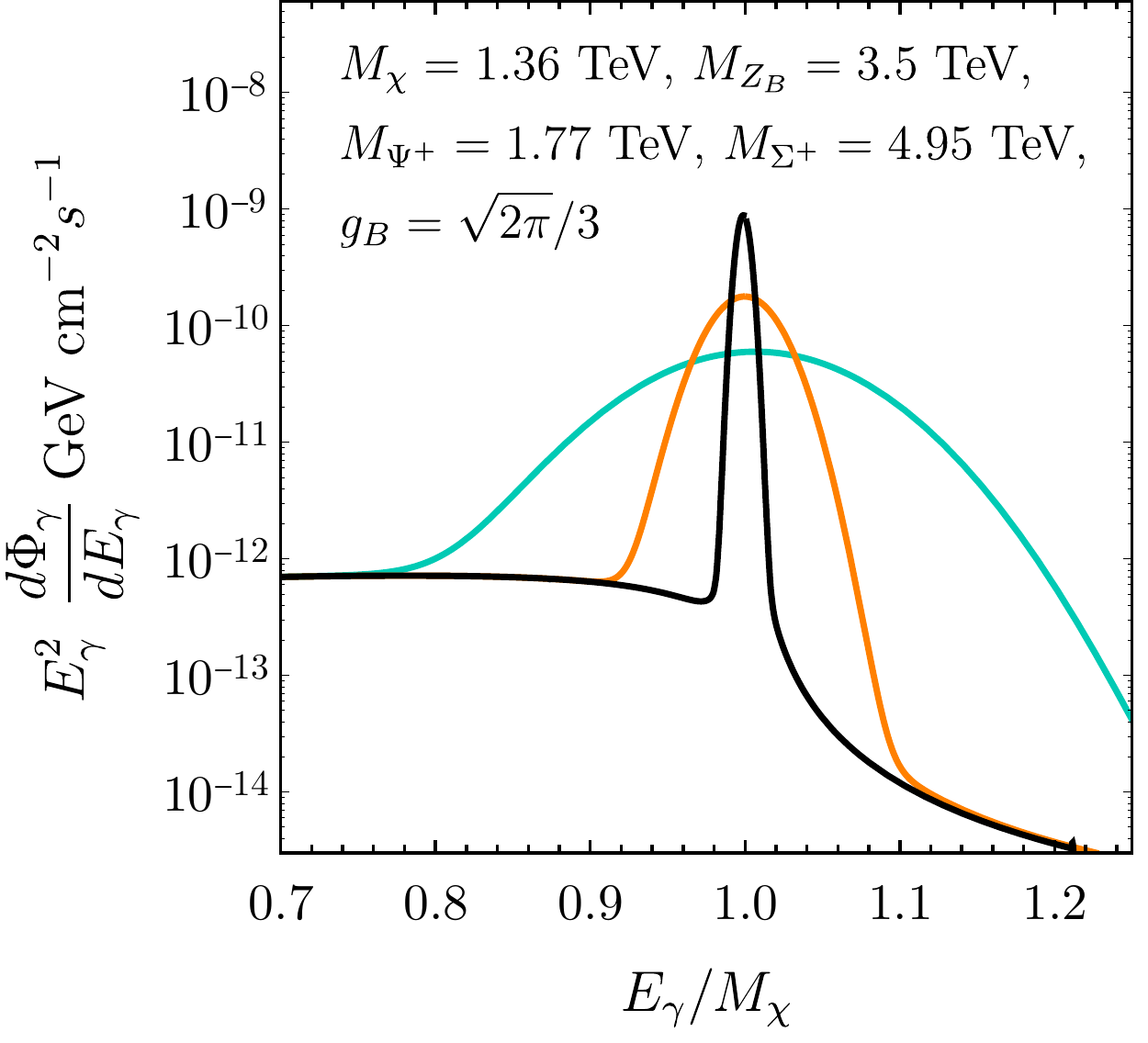} 
\hspace{1.3\textwidth} \includegraphics[width=0.4\linewidth]{leg.pdf} 
\caption{The differential spectrum of the dark matter annihilation into gamma rays for two different scenarios that satisfy the relic abundance measured by the Planck satellite $\Omega_{\rm{DM}} h^2 = 0.1197 \pm 0.0022$ \cite{Ade:2015xua} in Model II. We set $M_{h_2}=500$ GeV and $\theta_B=0$. Lines with different colors correspond to different energy resolutions $\xi = \{ 0.01, 0.05, 0.15\}$.}
\label{GLsignalsII}
\end{figure}
%%%%%%%%%%%%%%%%%%%%%%%%%%%%%%%%%%%%%%%%%%%
%%%%%%%%%%%%%%%%%%%%%%%%%%
\subsubsection{Gamma Line Spectra}
%%%%%%%%%%%%%%%%%%%%%%%%%%%%%%%%
The prediction for the gamma-ray flux is given by, 
\begin{equation}
\frac{d\Phi_\gamma}{dE_\gamma}= \frac{n_\gamma }{8 \pi M_\chi^2} \frac{ d\langle \sigma v_\text{rel} \rangle}{dE_\gamma}J_\text{ann}
=\frac{n_\gamma \langle \sigma v_\text{rel} \rangle}{8\pi M_\chi^2}\frac{dN}{dE_\gamma} J_\text{ann},
\end{equation}
%
% Our predictions for the annihilation cross-section into two photons and a photon and a $Z$ boson are presented in Figs.~\ref{} and~\ref{} for Model I and II respectively. 
where $n_\gamma$ is a multiplicity factor, $n_\gamma=2 \, \, (=1)$ for the $\gamma \gamma \, \, (\gamma Z)$ annihilation channel. In order to compute the total flux, we include the gamma lines cross-section and the final state radiation. The $J$ factor, $J_\text{ann}$, encodes all astrophysical assumptions made regarding the dark matter distribution. 
Here, we will use the value $J_\text{ann}=13.9 \times 10^{22} \text{ GeV}^2 \text{cm}^{-5}$~\cite{Ackermann:2015lka,Ackermann:2013uma} for our numerical analysis. The spectrum function is given by
\begin{equation}
\frac{dN}{dE_\gamma}=\int_0^\infty dE_0 \,  W_\text{final} \, G(E_\gamma, \xi /\omega, E_0),
\end{equation}
where $W_\text{final}$ is $W_{\gamma \gamma} = \delta(E_0 - M_\chi)$ for the annihilation into two photons and
\begin{equation}
W_{\gamma Z} = \frac{1}{\pi}\frac{4M_\chi M_Z \Gamma_Z}{(4M_\chi^2-4M_\chi E_0 - M_Z^2)^2+4 \Gamma_{Z}^2M_Z^2},
\end{equation}
for the $Z \gamma$ line. In order to account for the energy resolution of the detector, we apply a Gaussian function,
%Here, we use a Gaussian function to model the detector resolution, $G(E_\gamma,\xi/\omega, E_0)$, which reads as
%
\begin{equation}
G(E_\gamma,\xi/\omega, E_0)=\frac{1}{\sqrt{2\pi}E_0(\xi/\omega)} e^{ -\frac{(E_\gamma - E_0)^2}{2E_0^2(\xi/\omega)^2}},
\end{equation}
where $\xi$ is the energy resolution and  $\omega = 2 \sqrt{2 \rm{log} 2} \approx 2.35$ determines the full width at half maximum, with the standard deviation given by $\sigma_0=E_0 \xi/w$.

In Fig.~\ref{GLsignalsI} we present our results for the differential spectrum of dark matter annihilation into gamma rays for parameters satisfying the relic abundance, $\Omega_{\rm{DM}} h^2 = 0.1197 \pm 0.0022$, in Model I and different values of the energy resolution $\xi = \{ 0.01, 0.05, 0.15\}$. 
As can be appreciated, the gamma lines can be easily distinguished from the continuum spectrum because the FSR processes are highly suppressed.
However, for gamma ray telescopes with energy resolution larger than $5 \%$ it is impossible to distinguish the gamma lines between the $\chi \chi \to \gamma \gamma$ and $\chi \chi \to \gamma Z$ annihilation channels. In Fig.~\ref{GLsignalsII} we show 
the predictions for the differential spectrum of the dark matter annihilation into gamma rays for different values saturating the relic abundance in two different scenarios in Model II. 
The predictions for the gamma lines in Models I and II are very striking but very similar, and hence, collider searches are needed in order to distinguish between the two models. This is the goal of our future publication.

%%%%%%%%%%%%%%%%%%%%%%%%%%%%
\section{SUMMARY}
\label{sec:Summary}
%%%%%%%%%%%%%%%%%%%%%%%%%%%%
In this work, we investigated the properties of a Majorana dark matter candidate predicted in anomaly-free theories.
In these theories, the dark matter mass is defined by the new symmetry breaking scale and its stability is ensured by a remnant discrete symmetry after the gauge symmetry is broken.
We focus our study in the context of simple theories where the baryon number is a local gauge symmetry spontaneously broken at the low scale. 
Our dark matter candidate is a Majorana fermion charged under the $\U(1)_B$ gauge group, and hence, it has an axial coupling to the new gauge boson $Z_B$. 
We performed the calculation of the relic density and discussed experimental constraints coming from the LHC and direct detection experiments. The axial coupling implies that for vanishing Higgs mixing the dark matter-nucleon cross-section is velocity-suppressed. Therefore, the bounds from direct detection experiments only become relevant for large Higgs mixing. However, dijet resonance searches at the LHC will further probe the regions in which $M_{Z_B}$ lies around the electroweak scale. 
Moreover, we found that the cosmological constraint on the dark matter relic density, $\Omega_{\rm{DM}} h^2 \leq 0.12$, implies that the new gauge boson and the dark matter masses must be below the multi-TeV scale, i.e. $M_{Z_B} \lesssim 28 \  {\rm TeV}$ and $M_\chi \lesssim 34 \  {\rm TeV}$.

The new fermions needed for anomaly cancellation in these theories are chiral under the $\U(1)_B$ and acquire their masses after this symmetry has been spontaneously broken. 
Therefore, the upper bound on the baryon number violation scale translates as an upper bound on their masses. These upper bounds tell us that there is hope to 
detect this new sector of the theory at the LHC or at future particle colliders.
Regarding indirect detection, the dark matter axial coupling implies that the final state radiation is velocity suppressed. Consequently, the gamma lines from dark matter annihilation can be distinguished from the continuum.
We would like to emphasize that consistent completions of simplified models of dark matter 
give rise to interesting phenomenology; namely, gamma line features that can be probed at future gamma ray telescopes.  Due to the gauge anomaly conditions, 
these signatures cannot be computed within a simplified model where the anomalies are not canceled. Our predictions for the gamma lines could be crucial to 
test these dark matter theories in the future.\\

{\small{{\textit{Acknowledgments:}}  
The work of C.M. has been supported in part by Grants No. FPA2014-53631-C2-1-P, FPA2017-84445-P and SEV-2014-
0398 (AEI/ERDF, EU), and La Caixa-Severo Ochoa scholarship. }}
%%%%%%%%%%%%%%%%%%%%%%%%%%%%%%%%%%%%%%%%%%%%%%%%%%%%%%%%%%%%%%%%%%%%%%%%%%%%%%%

%%%%%%%%%%%%%%%%%%%%%%%%%%%%%%%%%%%%%%%%%%%%%%%%%%%%%%%%%%%%%%%%%%%%%%%%%%%%%%%
%%%%%%%%%%%%%%%%%%%%%%%%%%%%%%%%%%%%%%%%%%%%%%%%%%%%%%%%%%%%%%%%%%%%%%%%%%%%%%%
%%%%%%%%%%%%%%%%%%%%%%%%%%%%%%%%%%%%%%%%%%%%%%%%%%%%%%%%%%%%%%%%%%%%%%%%%%%%%%
%%%%%%%%%%%%%%%%%%%%%%%%%%%%%%%%%%%%%%%%%%%%%%%%%%%%%%%%%%%%%%%%%%%%%%%%%%%%%%%
%
%%%%%%%%%%%%%%%%%%%%%%%%%%%%%%%%%%%%%%%%%%%%%%%%%%%%%%%%%%%%%%%%%%%%%%%%%%%%%%%
%%%%%%%%%%%%%%%%%%%%%%%%%%%%%%
%

%%%%%%%%%%%%%%%%%%%%%%%%%%%%%%%%%%%%%%%%%%%%%%%%%%%%%%
% BIBLIOGRAPHY
\bibliography{MajoranaLeptophobic}
%%%%%%%%%%%%%%%%%%%%%%%%%%%%%%%%%%%%%%%%%%%%%%%%%%%%%%

\end{document}